\documentclass[a4paper,12pt]{article} 
\usepackage[T1]{fontenc}
\usepackage{amsfonts,amsmath,amssymb,enumerate,epsfig,graphicx}
\usepackage{appendix} 
\usepackage{anysize}
\usepackage{makeidx}
\usepackage{url}
\usepackage{tikz}
\usepackage{color}
\usepackage[colorlinks=false,linkcolor=red,citecolor=green,urlcolor=green,linktocpage=true,hyperindex=true]{hyperref}

\numberwithin{equation}{section}

\newtheorem{definition}{Definition}[section]
\newtheorem{proposition}[definition]{Proposition}
\newtheorem{remark}[definition]{Remark}
\newtheorem{theorem}[definition]{Theorem}
\newtheorem{lemma}[definition]{Lemma}

\newcommand{\prf}{\underline{Proof:}\ }
\newcommand{\finprf}{\null \hfill {\rule{5pt}{5pt}}\\ \null}

\newcommand{\mm}{\mathbf{m} }

\newcommand{\qq}{\mathbf{q} }
\newcommand{\pp}{\mathbf{p} }
\newcommand{\rp}{\mbox{$ \breve{\mathbf{p} }  $}}
\newcommand{\vv}{\mathbf{v} }
\newcommand{\uu}{\mathbf{u} }

\newcommand{\be}{\begin{equation}}
\newcommand{\ee}{\end{equation}} 
\newcommand{\bma}{\begin{pmatrix}}
\newcommand{\ema}{\end{pmatrix}}
\newcommand{\bea}{\begin{eqnarray}}
\newcommand{\eea}{\end{eqnarray}}

\newcommand{\KE}{\operatorname{Ker}}
\newcommand{\IM}{\operatorname{Im}}

\newcommand{\RR}{\mbox{${\mathbb R}$}}
\newcommand{\CC}{\mbox{${\mathbb C}$}}
\newcommand{\PP}{\mbox{${\mathbb P}$}}
\newcommand{\yR}{\mbox{$ \cal R  $}}
\newcommand{\rR}{\mbox{$ \cal B  $}}

\makeatletter

\let\captionORI\caption
\def\caption#1{\captionORI{\rm\small #1}}
\makeatother

\begin{document}

\pagestyle{empty}
\setcounter{page}{0}
\null

\begin{center}

{\LARGE \bf  Yang-Baxter and reflection maps\\
from vector solitons with a boundary}

\vspace{1cm}

{\large V. Caudrelier and Q. C. Zhang }\\

\vspace{0.5cm}

{\it Department of Mathematics,
City University London, \\
Northampton Square,London EC1V 0HB, UK.}
\end{center}

\vfill

\begin{abstract}
Based on recent results obtained by the authors on the inverse scattering method of the vector nonlinear Schr\"odinger equation with integrable boundary 
conditions, we discuss the factorization of the interactions of $N$-soliton solutions on the half-line. Using dressing transformations combined with a mirror image technique, factorization of soliton-soliton and soliton-boundary interactions is proved. 
We discover a new object, which we call 
reflection map, that satisfies a set-theoretical reflection equation which we also introduce. Two classes of solutions for the reflection map are constructed.
Finally, basic aspects of the theory of the set-theoretical reflection equation are introduced.
\end{abstract}


AMS classification (2010): 35Q55, 16T25

\vspace{4cm}

\hspace{8cm} \textbf{\it Nonlinearity 27 (2014) 1081-1103. }

\hspace{8cm} \textbf{\it Recommended by A S Fokas.}

\vfill

\newpage
\pagestyle{plain}
\section{Introduction}\label{sec:1}

In this article, we consider the Manakov model \cite{1974JETP...38..248M} or more generally the vector nonlinear Schr\"odinger equation (VNLS) 
for the n-component complex-valued vector field $R(x,t)$, in the focusing case,
\begin{align} 
  \label{equ:schr}
  i \frac{\partial R}{\partial t} + \frac{ \partial^2 R}{\partial x ^2} + 2  R R^\dagger R = 0\,,\\
  \end{align}
where  $R^\dagger(x,t)$ is the transpose  conjugate of $R(x,t)$, and we restrict it to the half-line $x\ge 0$ by imposing 
integrable boundary conditions derived in \cite{2011arXiv1110.2990C}: Robin boundary conditions
\bea
\label{Robin_BC}
R_x(0,t)-2\alpha R(0,t)=0~~,~~\alpha\in\RR\,,
\eea
or a mixture or Neumann and Dirichlet boundary conditions
\bea
\label{mixed_BC1}
&&R_j(0,t)=0~~,~~j\in S~~,\\
\label{mixed_BC2}
&&R_{kx}(0,t)=0~~,~~k\in \{1,\dots,n\}\setminus S\,,
\eea
where $S$ is a subset of $\{1,\dots,n\}$. Supplemented with the initial condition $R(x,0)=R_0(x)$, we therefore consider an integrable initial-boundary value 
problem\footnote{Of course, initial and boundary values are supposed to be compatible and, to fix ideas, we work in the space of 
Schwarz functions which is known to be appropriate for the inverse scattering method, see e.g. \cite{faddeev2007hamiltonian}}.
In \cite{2011arXiv1110.2990C}, the inverse scattering method valid for the problem on the full line has been suitably adapted to yield 
reconstruction formulae for the function $R(x,t)$. In particular, explicit $N$-soliton solutions have been obtained. An interesting effect 
has been identified when a soliton bounces off the boundary: its polarization can be altered and a redistribution of the amplitudes 
of the various components occurs through a process of transmission between modes. However, a full understanding of the dynamics 
of the solitons on the half-line is still lacking and it is the object of this paper to discuss this point.

For the problem on the line, this question 
has only been studied rather recently in \cite{2004PThPh.111..151T,0266-5611-20-4-012}. The fundamental result is that soliton interactions (or collisions) 
factorize in the vector case like in the scalar case. 
In the scalar case, factorization of soliton interactions is a well-known and important fact. Roughly speaking, it means that as time evolves from 
$-\infty$ to $\infty$, two solitons with given (different) velocities and amplitudes as $t\to - \infty$ will collide and recover their initial properties 
(velocities, amplitudes, shapes), the effect of the interaction being only position and phase shifts. This fact generalizes to $N$-soliton collisions. 
In this case, 
the resulting velocities and amplitudes of the solitons as $t\to  \infty$ is independent of the way the collisions took places. This
 gives rise to the notion of factorization of soliton interactions. In the vector case, solitons are specified by their so-called polarization vectors 
 on top of their velocities and amplitudes, and the property of factorization now involves these polarizations. 
 It is therefore a highly nontrivial fact that factorization still holds. In terms of the polarizations, this means that the set of 
polarization vectors of the solitons obtained as $t\to\infty$ is independent of the way the collisions between solitons took place but depends only on the 
 initial polarization vectors. 
 
 This property in the vector case is intimately related to the existence of a so-called Yang-Baxter map 
 which satisfies the set-theoretical Yang-Baxter equation (YBE) as originally proposed in \cite{springerlink:10.1007/BFb0101175} and studied in 
 more details in \cite{weinstein1992classical,etingof-1997,lu2000set}. 
 This map describes the relation between two polarization vectors before and after the collision. 
 The proof of the factorization property was established using two related but rather different routes in \cite{2004PThPh.111..151T} and 
 \cite{0266-5611-20-4-012}. In \cite{2004PThPh.111..151T}, the final result of the inverse scattering method, i.e. the explicit $N$-soliton 
 solution, is carefully studied in the limits $t\to\pm\infty$ to extract the factorization property by looking directly at the 
 evolution of the polarization vectors. In \cite{0266-5611-20-4-012}, the results obtained originally by Manakov by asymptotic analysis of the 
 inverse scattering method are used to construct an inductive argument on the number of solitons. Then, this is cast in the formalism 
 of Yang-Baxter maps and the associated Lax pairs to confirm factorization using matrix polynomials refactorization properties. 
 The Yang-Baxter map is realised in this context by a map acting on pairs of polarization vectors. It is the mathematical representation of 
 a soliton-soliton interaction.
 
In essence, the set-theoretical YBE ensures the factorization property. The situation is analogous to the 
quantum case where factorization is ensured by the quantum YBE but we emphasize that the nature of the objects that 
are involved is fundamentally different. In fact, the set-theoretical YBE involves maps on cartesian products of sets and 
provides a very general framework. The quantum YBE can be seen as a special case of general theory.

By studying the question of factorization in the presence of a boundary, we discover a new framework which, 
jointly with the set-theoretical YBE, ensures the factorization property on the half-line: the \textit{set-theoretical reflection equation}. 
We also find two classes of solutions for this equation which we call \textit{reflection maps}.
 
 The paper is organised as follows. In section \ref{sec:2}, we use the formalism of the Riemann-Hilbert problem and the dressing method 
 \cite{springerlink:10.1007/BF01075696,springerlink:10.1007/BF01077483} to 
 establish Theorem \ref{prop23} which is the central technical tool in our approach. First, it allows us to rederive the above factorization property on the line 
 directly at the level of the dressing method. On top of providing a fresh look and a new proof of the result, this allows us to introduce all the relevant definitions, notations and results needed for 
 our purposes. Then, in section \ref{sec:boundarycond}, we combine these results with the mirror image technique developed in \cite{2011arXiv1110.2990C} to prove the factorization of 
 the soliton-soliton \textit{and} soliton-boundary interactions. This gives rise to the introduction of a new object, \textit{the reflection map}, 
 which is a map acting on one polarization vector at a time and is the mathematical representation of the soliton-boundary interaction. 
 Just like the factorization of soliton-soliton interactions is related to the set-theoretical YBE, we find that the factorization property
 with a boundary is related to an equation which we call the \textit{set-theoretical reflection equation}. This and the construction of 
 explicit reflection maps form the main result of this paper presented in Theorem \ref{prop:refmap2}. Section \ref{RM} is then devoted to the presentation of a 
 few basic elements of the theory of set-theoretical RE in an abstract setting. 
A few technical arguments are collected in appendices.

It is remarkable that our results complete the formal analogy that exists between quantum integrable systems and classical soliton partial differential equations.
We already mentioned this analogy at the level of the YBE. Here, our findings provide a further link with the reflection equation that appears in the theory of quantum integrable 
systems with boundaries \cite{springerlink:10.1007/BF01038545,sklyanin1988boundary}. In the particular case of VNLS, the complete classification of the solutions of the 
quantum reflection equation was obtained in \cite{mintchev2001spontaneous}. 

\section{Dressing method, factorization and Yang-Baxter maps}\label{sec:2}

In this section, we first  review some basic concepts of the inverse scattering method (ISM) and its relation to the  dressing method. 
The main contribution  relies on the permutability property of dressing transformations  which is stated in Theorem \ref{prop23}. It then 
allows us to provide a new proof of the factorization property for VNLS on the line and to show the underlying Yang-Baxter structure.

We refer readers to e.g. \cite{faddeev2007hamiltonian, ablowitz2004discrete, 2006nlin......4004G} for more detailed presentations of the ISM, 
and  to e.g.  \cite{intro_2_CIS} for a presentation of the dressing method. 

\subsection{Lax pair formulation}
\label{sec:21}
Define $Q(x,t)$ as the following $(n+1) \times (n+1)$ matrix-valued field
\begin{equation} \label{equ:field}
Q(x,t) = \bma 0 & R(x,t) \\ - R^{\dagger}(x,t) & 0 \ema\,. 
\end{equation}
The VNLS equation \eqref{equ:schr} is the compatibility condition ($\Phi_{xt}= \Phi_{tx}$) of the following linear problems for the matrix-valued function 
$\Phi(x,t,k)$\footnote{From now on, we drop the $x$, $t$ and $k$ dependence for conciseness unless there is ambiguity.} 
\begin{align}
  \label{equ:laxpx} 
&\Phi_x + ik [ \Sigma _3 , \Phi] =  Q\, \Phi \,,  \\ 
\label{equ:laxpt} 
&\Phi_t + 2 i k^2 [\Sigma_3, \Phi] =   Q_T\, \Phi\,,  
\end{align}
where
\begin{equation}
  \label{eq:qt}
 \Sigma_3 = \bma I_n & 0 \\ 0 & -1  \ema\,, \quad  Q_T = 2k Q -  i Q_x \, \Sigma_3 -  iQ^2 \, \Sigma_3 \,, 
\end{equation}
$I_n$ being the $n\times n$ identity matrix.  Equations (\ref{equ:laxpx},  \ref{equ:laxpt}) form the Lax pair formulation for VNLS. 
One observes that  $Q$ satisfies
\begin{equation}
  \label{eq:laxpair1}
  Q \,  =  - Q^\dagger\,,
\end{equation}
where $^\dagger$ denote the transpose conjugate operation. This implies that, $\Phi(x,t,k)$ being solution of the Lax pair, $\Phi^\dagger(x,t,k^*) $ 
satisfies the same equations as 
$\Psi(x,t,k) \equiv \Phi^{-1}(x,t,k)$ i.e.
\begin{align}
  \label{equ:laxpx1} 
&\Psi_x + ik [ \Sigma _3 , \Psi] =  - \Psi\,Q \,,  \\ 
\label{equ:laxpt1} 
&\Psi_t + 2 i k^2 [\Sigma_3, \Psi] =   - \Psi\, Q_T\,.
\end{align}
We define two Jost  solutions $X$ and $Y$ of the Lax pair (\ref{equ:laxpx}, \ref{equ:laxpt}) satisfying 
\begin{align}
\label{eq:asympx}
 \lim_{x\to -\infty}e^{i \phi(x,t,k) \,  \Sigma_3 }   X(x,t,k) e^{- i \phi(x,t,k) \,  \Sigma_3 }= &I_{n+1}\, , \quad k \in \RR \,,\\ 
\label{eq:asympy}
 \lim_{x\to \infty}e^{i \phi(x,t,k) \,  \Sigma_3 }  Y(x,t,k) e^{- i \phi(x,t,k) \,  \Sigma_3 } = &I_{n+1}\, ,\quad  k \in \RR\,,
\end{align}
where $\phi(x,t,k) =  kx + 2k^2t$.  They enjoy the following properties
\begin{itemize}

\item 
\begin{equation}
  \label{eq:det1}
  \det X(x,t,k) = \det Y(x,t,k ) = 1\,.
\end{equation}

\item $X$ and $Y$ can be splitted into the following "column" vectors\footnote{Here, the left "column" vector is made of the first left $n$ columns and 
the right one is made of the remaining column. This column vector representation will always be used in the rest of this paper.} form
\be
\label{eq:xy11}
 X =(X^+,X^-) \,, \quad Y = (Y^-,Y^+) \,,
\ee
where  $ X^+$, $ Y^+$ (resp. $X^-$, $Y^-$) are analytic and bounded in the upper (resp. lower) half $k$-complex plane, which are denoted by $\CC^+$ (
resp. $\CC^-$).  

\item $X$ and $Y$ are related by the so-called scattering matrix $S(k)$   
\be
\label{def_S}
 X(x,t,k) = Y(x,t,k) e^{-i\phi(x,t,k)\Sigma_3}\, S(k) \,e^{i\phi(x,t,k)\Sigma_3}\,, \quad k \in \RR   \,,
\ee
where $S(k)$ can be splitted into block matrices of natural sizes\footnote{For instance, $a^+$ is an $n\times n$ matrix while $a^-$  is a scalar.} 
\bea\label{rep_S2}
S(k) =  \begin{pmatrix}
a^+(k)&b^-(k)\\
b^+(k)&a^-(k)
\end{pmatrix} \,,
\eea 
where $a^\pm(k)$ allows for analytic continuation into $\CC^\pm$ respectively. 

\item 
\begin{equation}
\label{reduction1}
X^{-1}(x,t,k) = X^\dagger(x,t,k^*)~~,~~Y^{-1}(x,t,k) = Y^\dagger(x,t,k^*)\,,
\end{equation}
and hence 
\begin{equation}
\label{reduction2}
S(k)^{-1} = S^\dagger(k^*)\,.
\end{equation}
In components, we denote
\bea\label{rep_T2} 
S(k)^{-1} =  \begin{pmatrix}
c^-(k)&d^-(k)\\
d^+(k)&c^+(k)
\end{pmatrix} \,.
\eea 
Accordingly, $c^\mp(k)$  has an analytic continuation into $\CC^\mp$. In particular, one has the following relations
 \begin{eqnarray}
  \label{eq:det111}
  (a^\pm) ^\dagger(k^*)  = c^{\mp}(k)  \,, \quad  k \in \CC^\mp\,,\\
  \label{eq:det112}
  \det a^+(k) = c^+(k)  \,, \quad  k \in \CC^+\,.
\end{eqnarray}
\end{itemize}
For VNLS, there are two equivalent sets  of scattering functions $\{a^\pm(k), b^\pm(k)\}$ and $\{c^\pm(k), d^\pm(k)\}$ 
available to reconstruct $R(x,t)$ in the inverse part of the ISM \cite{2006nlin......4004G}.
In the rest of this paper, we choose to work with $a^\pm(k)$ and $b^{\pm}(k)$. The information on the soliton solutions depends on the analytic structure 
of $a^\pm(k)$.

A key observation in the development of soliton theory is that the ISM can be cast into a Riemann-Hilbert problem. The latter offers a natural and powerful 
framework for the so-called dressing method \cite{springerlink:10.1007/BF01075696,springerlink:10.1007/BF01077483}. In the next subsection, we present general results about 
the dressing method before returning to the case of VNLS.

\subsection{Riemann-Hilbert problem and dressing method }\label{sec:RH11}

We mainly follow the presentation of chapter $3$ of \cite{intro_2_CIS} and begin by stating the main propositions about the connection between Riemann-Hilbert (RH) problems with zeroes and 
the dressing method. We refer readers to \cite{intro_2_CIS}  for  details and in particular for the proofs of Propositions \ref{prop:dressing15} and \ref{prop:dressing17} below. 

Consider the following matrix RH problem with canonical normalization
\begin{equation}
  \label{eq:rrrh222}
  {\cal J}^+(k) {\cal J}^- (k)   =    {\cal J}(k)\,,\quad k \in \RR\,, \quad \lim_{|k|\to \infty} {\cal J}^\pm(k) \to I \, ,
\end{equation} 
where ${\cal J}(k)$ is the jump matrix satisfying $\det {\cal J}(k)\neq 0$ for $k\in\RR$.
This RH problem is discussed in some details for instance in section $7.5$ of \cite{Ablowitz:1138676}. The matrix ${\cal J}^\pm(k)$ is  analytic in $\CC^\pm$.
This problem has a unique regular solution ${\cal J}_0^\pm(k)$ that is a solution with 
$\det {\cal J}_0^\pm(k)\neq 0$. 
In the construction of soliton solutions, the notion of RH problems with zeroes plays an important role. Although not the most general 
one (see e.g. \cite{shchesnovich2003general}), the following definition is sufficient for our purposes.
\begin{definition}
A matrix $M(k)$ is said singular at $k=k_0$ if $\det M(k_0)=0$ and if in the neighbourhood of $k_0$
\begin{eqnarray}
M(k)=M_0+(k-k_0)M_1+O(k-k_0)^2\,,\quad M^{-1}(k)=\frac{N_0}{k-k_0}+N_1+O(k-k_0)\,.
\end{eqnarray}
\end{definition}
\begin{definition}
A RH problem with zeroes at $k_j^\pm\in\CC^\pm$, $j=1,\dots,N$ is a RH problem as in \eqref{eq:rrrh222} where ${\cal J}^\pm(k)$ is singular at $k_j^\pm$, $j=1,\dots,N$.
\end{definition}
Then one proves
\begin{proposition}
\label{prop:dressing15}
Fixing the subspaces ${\cal V}_j \equiv \left. \IM \, {\cal J}^+ (k)\right|_{k= k^+_j}$ and ${\cal U}_j \equiv \left. \KE \, {\cal J}^- (k)\right|_{k= k^-_j}$, $j=1,\dots,N$ 
determines uniquely 
the solution of the RH problem with zeroes at $k_j^\pm\in\CC^\pm$.
\end{proposition}

In general, there is no known closed-form formula to solve a matrix RH problem. However, once a regular solution is known, it is possible to construct singular solutions from it.
\begin{proposition}
  \label{prop:dressing17}
Let ${\cal J}^\pm(k)$ be the singular solution at $k^\pm_0\in \CC^\pm$ with
\begin{equation}
  \label{eq:dressing18}
  \left. \IM {\cal J}^+(k) \right|_{k = k^+_0} = {\cal V}_0 \, , \quad   \left. \KE {\cal J}^-(k) \right|_{k = k^-_0}  = {\cal U}_0\,, 
\end{equation}
and let ${\cal J}_0^\pm(k)$ be the solution of the same RH problem regular at $k_0^\pm$. Then $ {\cal J}^\pm(k)$ can be written as  
\begin{equation}
\displaystyle
\label{eq:dressing40}
  {\cal J}^+(k)  =  {\cal J}_0^+ (k)(I_{n+1 }  +  \frac{k_0^--k^+_0}{k-k^-_0} \Pi_0) \,, \quad  
  {\cal J}^-(k) =   (I_{n+1 }  +  \frac{k^+_0-k^-_0}{k-k^+_0} \Pi_0) {\cal J}_0^-(k) \,.  
\end{equation}
Here $\Pi_0$ is the projector defined by
\begin{equation}
  \label{eq:dressing21}
  \KE \Pi_0   = \left({\cal J}_0^+(k^+_0)\right)^{-1}{\cal V}_0\,, \quad   \IM \Pi_0   = {\cal J}_0^-(k^-_0){\cal U}_0 \,.
\end{equation}
\end{proposition}
Proposition \ref{prop:dressing17} introduces what is called a dressing factor (of degree $1$) which transforms  ${\cal J}_0^\pm(k)$ regular at $k_0^\pm$ into ${\cal J}^\pm(k)$ 
singular at $k^\pm_0$. This gives an algorithm to construct a singular solution ${\cal J}^\pm(k)$ at \textit{distinct} $ k^\pm_j \in \CC^\pm$, $j = 1 ,\dots , N$
from a regular solution ${\cal J}^\pm_0(k)$. Precisely,  let $k_j^\pm\in\CC^\pm$, $j = 1, \dots , N$ and the 
corresponding subspaces ${\cal V}_j$, ${\cal U}_j$ be given. Use Proposition \ref{prop:dressing17} repeatedly to construct 
${\cal J}^\pm(k)$ singular at $k_j^\pm$ recursively from ${\cal J}^\pm_0(k)$ starting from $k_1^\pm$, $k_2^\pm$ up to $k_N^\pm$.
Consequently, ${\cal J}^\pm(k)$ can be written as
\begin{align}
\label{original_dressing1}
  {\cal J}^+(k) =& {\cal J}_0^+ (k)\left(I_{n+1} + \frac{k_1^- - k_1^+}{k-k_1^-} \Pi_1 \right) \dots  
  \left(I_{n+1} +     \frac{k_N^- - k_N^+}{k-k_N^-} \Pi_N \right) \,, \\
  \label{original_dressing2}
  {\cal J}^- (k)=&  \left(I_{n+1} + \frac{k_N^+ - k_N^-}{k-k_N^+} \Pi_N \right) \dots  
  \left(I_{n+1} +     \frac{k_1^+ - k_1^-}{k-k_1^+} \Pi_1 \right)  {\cal J}_0^-(k)\,,
\end{align}
where, for $j=1,\dots,N$
\begin{align}
  \label{eq:dres2111}
    \KE \Pi_j   =   & \left(  {\cal J}^+_0(k^+_j) \left(I_{n+1} + \frac{k_{1}^- - k_{1}^+}{k^+_j-k_{1}^-} \Pi_{j-1} \right) \dots  
    \left(I_{n+1} +     \frac{k_{j-1}^- - k_{j-1}^+}{k^+_{j} -k_{j-1}^-} \Pi_1 \right)\right)^{-1}{\cal V}_j\,, \\  
    \IM \Pi_{j}  = &  \left(I_{n+1} + \frac{k_{j-1}^+ - k_{j-1}^-}{k^-_j-k_{j-1}^+} \Pi_{j-1} \right) \dots  
    \left(I_{n+1} +     \frac{k_1^+ - k_1^-}{k^-_{j} -k_1^+} \Pi_1 \right) {\cal J}^-_0(k^-_j)\, {\cal U}_j \,.
\end{align}
Now comes a simple but fundamental observation which is absent in chapter $3$ of \cite{intro_2_CIS}. In the above construction, one can iterate Proposition \ref{prop:dressing17} by using 
a different order on the $k_j^\pm$. Let $S_N$ be the permutation group on the set $\{1,\dots,N\}$ and let $\sigma\in S_N$. Denote 
the image of $(1,\dots,N)$ under $\sigma$ by $(\sigma(1),\dots,\sigma(N))$ and introduce $\kappa^\pm_j=k^\pm_{\sigma(j)}$. 
Then, the subspaces corresponding to $\kappa_j^\pm$ are ${\cal V}_{\sigma(j)}$, ${\cal U}_{\sigma(j)}$. Repeating the previous procedure, starting from 
$\kappa_1^\pm$ up to $\kappa_N^\pm$, we obtain
\begin{align}
\label{permuted_dressing1}
\tilde{{\cal J}}^+(k) =& {\cal J}_0^+ (k)\left(I_{n+1} + \frac{k_{\sigma(1)}^- - k_{\sigma(1)}^+}{k-k_{\sigma(1)}^-} \Pi^\sigma_1 \right) \dots  
  \left(I_{n+1} +     \frac{k_{\sigma(N)}^- - k_{\sigma(N)}^+}{k-k_{\sigma(N)}^-} \Pi^\sigma_N \right) \,, \\
  \label{permuted_dressing2}
  \tilde{{\cal J}}^- (k)=&  \left(I_{n+1} + \frac{k_{\sigma(N)}^+ - k_{\sigma(N)}^-}{k-k_{\sigma(N)}^+} \Pi^\sigma_N \right) \dots  
  \left(I_{n+1} +     \frac{k_{\sigma(1)}^+ - k_{\sigma(1)}^-}{k-k_{\sigma(1)}^+} \Pi^\sigma_1 \right)  {\cal J}_0^-(k)\,,
\end{align}
where, for $j=1,\dots,N$
\begin{align}
    \KE \Pi^\sigma_j   =   & \left({\cal J}^+_0(k^+_{\sigma(j)})  \left(I_{n+1} + \frac{k_{\sigma(1)}^- - k_{\sigma(1)}^+}{k^+_{\sigma(j)}-k_{\sigma(1)}^-} \Pi^\sigma_{j-1} \right) 
    \dots  \left(I_{n+1} +     \frac{k_{\sigma(j-1)}^- - k_{\sigma(j-1)}^+}{k^+_{\sigma(j)} -k_{\sigma(j-1)}^-} \Pi^\sigma_1 \right) 
    \right)^{-1}{\cal V}_{\sigma(j)}\,, \\  
    \IM \Pi^\sigma_{j}  = &  \left(I_{n+1} + \frac{k_{\sigma(j-1)}^+ - k_{\sigma(j-1)}^-}{k^-_{\sigma(j)}-k_{\sigma(j-1)}^+} \Pi^\sigma_{j-1} \right) \dots  
    \left(I_{n+1} +     \frac{k_{\sigma(1)}^+ - k_{\sigma(1)}^-}{k^-_{\sigma(j)} -k_{\sigma(1)}^+} \Pi^\sigma_1 \right) {\cal J}^-_0(k^-_{\sigma(j)}){\cal U}_{\sigma(j)} \,.
\end{align}
It can be checked by direct calculation that ${\cal V}_j = \left. \IM \, \tilde{{\cal J}}^+ (k)\right|_{k= k^+_j}$ and ${\cal U}_j = \left. \KE \, \tilde{{\cal J}}^- (k)\right|_{k= k^-_j}$, $j=1,\dots,N$
so that Proposition \ref{prop:dressing15} implies that $\tilde{{\cal J}}^\pm (k)={\cal J}^\pm (k)$. In turn, this implies that the product of dressing factors in \eqref{permuted_dressing1},
\eqref{permuted_dressing2} is equal to the product of dressing factors in \eqref{original_dressing1}, \eqref{original_dressing2}. 
This construction introduces the notion of dressing factor of degree $N$ which transforms $ {\cal J}^\pm_0(k)$ into $  {\cal J}^\pm (k)$ singular at $k_j^\pm$, $j = 1, \dots , N$. 
Proposition \ref{prop:dressing15} also ensures that a dressing factor of order $N$ factorises into $N$ dressing factors of degree $1$ and that the order of the factorization 
is irrelevant. At this stage, we note that this fact is known in various disguises (e.g. as the Bianchi permutativity property) and in various context (e.g. in the 
theory of matrix polynomials) but, to the best of our knowledge, it has not been presented anywhere in the above fashion. This is why we formulate it as a theorem below in a form 
suitable for our purposes.
\begin{remark}
It is important to realize that this does not mean at all that the individual factors in a dressing factor of degree $N$ commute. Indeed, in 
general $\Pi^\sigma_j\neq \Pi_{\sigma(j)}$. This can happen in special circumstances and in that case, the interaction of the solitons 
is trivial from the polarization point of view. The important message here is that, in the factorization of a dressing factor of degree $N$, 
the equations governing the projectors are crucial, in particular the order in which they appear is important. With this in mind, 
we introduce a notation that will help us formulate the main 
theorem of this section and prove the factorization of soliton interactions in the following sections.
\end{remark}

\begin{definition}
\label{def_general_dressing}
Given $ {\cal J}^\pm_0(k)$ a regular solution of the RH problem \eqref{eq:rrrh222}. Let $\sigma\in S_N$ be given and write $(\sigma(1),\dots,\sigma(N))=(i_1,\dots,i_N)$. 
Given $k^\pm_j$ and ${\cal V}_j$, ${\cal U}_j$, $j=1,\dots,N$, a general dressing factor of degree $1$ is defined as, for $1\le \ell\le N$,
\begin{equation}
D_{i_\ell,\{i_1 \dots i_{\ell-1}\}}(k)=I_{n+1} + \frac{k_{i_\ell}^- - k_{i_\ell}^+}{k -k_{i_\ell}^-} \Pi_{i_\ell,\{i_1 \dots i_{\ell-1}\}}
\end{equation}
where 
\begin{align}
  \KE \Pi_{i_\ell,\{i_1 \dots i_{\ell-1}\}}= & \left[D_{i_1}(k^+_{i_\ell})\dots D_{i_{\ell-1},\{i_1 \dots i_{\ell-2}\})}(k^+_{i_\ell}) \right]^{-1}
  \left({\cal J}^+_0(k^+_{i_\ell})\right)^{-1}{\cal V}_{i_\ell}\,,\\
  \IM \Pi_{i_\ell,\{i_1 \dots i_{\ell-1}\}}=  & \left[D_{i_1}(k^-_{i_\ell})\dots D_{i_{\ell-1},\{i_1 \dots i_{\ell-2}\}}(k^-_{i_\ell}) \right]^{-1}
  {\cal J}^-_0(k^-_{i_\ell}){\cal U}_{i_\ell}\,.
\end{align}
Finally, the dressing factor of degree $N$ is denoted as $D_{1\dots N}(k)$.
\end{definition}
Note that in the case $\ell=1$, we denote $D_{i_1,\{\}}(k)\equiv D_{i_1}(k)$. This convention will be adopted in the rest of the paper for any quantity involving sets of indices as subscripts.
With this definition and the understanding from the above discussion, we have proved the following 
\begin{theorem}
\label{prop23}
A dressing factor of degree $N$ can be decomposed into $N!$ equivalent products of $N$ dressing factors of degree $1$
\begin{equation}
  \label{eq:rh43}
  D_{1\dots N}(k) = D_{i_1}(k)\dots D_{i_{N},\{i_1 \dots i_{N-1}\}}(k)\,, 
\end{equation}
where $(i_1,\dots, i_N)$ is an arbitrary permutation of $(1, \dots, N)$.
\end{theorem}

In the next subsection we want to use the results of this subsection for a RH problem arising from a Lax pair formulation. The latter 
involves the variables $x$ and $t$ so that in fact, we are dealing with a parameter dependent RH problem of the form 
\begin{equation}
  \label{eq:rrrh22}
  {\cal J}^+(x,t,k) {\cal J}^- (x,t,k)   = {\cal J}(x,t,k)\,,\quad k \in \RR\,, \quad \lim_{|k|\to \infty} {\cal J}(x,t,k) \to I \, .
\end{equation} 
The success of the dressing method is related to the fact that all the results seen in this subsection go through for this parameter-dependent 
RH problem provided one works with $(x,t)$-dependent subspaces  ${\cal V}_j(x,t)$, ${\cal U}_j(x,t)$, $j=1,\dots,N$ (and the associated 
projectors) which are simply related to  ${\cal V}_j$, ${\cal U}_j$, $j=1,\dots,N$, by
\begin{eqnarray}
{\cal V}_j(x,t)= \varphi(x,t,k_j^+){\cal V}_j~~,~~{\cal U}_j(x,t)=\varphi(x,t,k_j^-){\cal U}_j\,,
\end{eqnarray}
where $\varphi$ is a solution of the so-called undressed Lax pair equations\footnote{Indeed, $\varphi$ satisfies  $U\varphi = \varphi_x$, $V\varphi = \varphi_t$, with  $U_t - V_x + [U,V] = 0$ 
(zero curvature condition). The Lax pair (\ref{equ:laxpx}, \ref{equ:laxpt}) is another way to express the zero curvature condition. For VNLS, $\varphi$ is simply related to $\Phi$ 
which satisfies (\ref{equ:laxpx}, \ref{equ:laxpt}) by $\varphi  = \Phi e^{-i \phi(x,t,k)\Sigma_3}$.  }.

\subsection{Application to VNLS: reduction}
\label{eq:reduction}

We begin by collecting some known facts that can be gathered for instance from \cite{faddeev2007hamiltonian, 2006nlin......4004G, Fokas:2008:UAB:1450929}.
The scattering system defined in \eqref{def_S} can be rewritten as the following $(x,t)$-dependent RH problem
\begin{equation}
  \label{eq:rh22}
  J^+(x,t,k) J^-(x,t,k)    =    J(x,t,k) \, , 
  \quad k \in \RR\,, \quad \lim_{|k|\to \infty} J^\pm(x,t,k) \to I_{n+1} \, .
\end{equation} 
This is achieved by defining
\begin{equation}\displaystyle
  \label{eq:rh0}
  J^+ (x,t,k)= \bma a^+(k) & 0 \\ 0 & c^+(k) \ema  \left( (X^+, Y^+\right)^{-1}(x,t,k) \, , \quad  J^-(x,t,k) =  \left( Y^-  , X^- \right) (x,t,k)\, , 
\end{equation}
and 
\begin{equation}
\label{eq:jump1}
  J(x,t,k) =e^{- i \phi(x,t,k) \Sigma_3 } \bma I & b^-(k) \\  d^+(k) & 1 \ema e^{i \phi(x,t,k) \Sigma_3 } \,, \quad k \in \RR\,.
\end{equation}
In particular,
\begin{equation}
  \label{eq:dressing11}
  \det J^+  (x,t,k)= \det a^+(k)\,, \quad  \det J^-(x,t,k) = a^-(k)\,.
\end{equation}
One then has
\begin{proposition}
\label{prop:rh1}
Consider the RH problem defined in \eqref{eq:rh22}. Let $J^\pm(x,t,k)$  be a solution of the Riemann-Hilbert problem, 
then  $J^+(x,t,k)$ (resp. $J^-(x,t,k)$ ) satisfies the Lax pair  (\ref{equ:laxpx1}, \ref{equ:laxpt1}) (resp. (\ref{equ:laxpx}, \ref{equ:laxpt})). In particular,  $J^+(x,t,k)$ gives a uniquely defined $Q(x,t)$ by 
\begin{equation}
  \label{eq:rhh1}
  Q(x,t) =  \lim_{|k|\to \infty} - ik [ \Sigma_3, J^+(x,t,k)]\,.
\end{equation}
\eqref{eq:rhh1} is called the reconstruction formula for $Q(x,t)$.  
\end{proposition}

Although in general one cannot solve the RH problem explicitely,  pure $N$-soliton solutions can be derived by using the dressing method. 
We follow the usual approach that corresponds to assuming that $\det a^+(k)$ (resp. $a^-(k)$) 
has a finite number $N$ of simple zeros $k^+_j \in \CC^+$ (resp. $k_j^- \in \CC^-$), $j = 1, \dots, N$. This means that we can 
use the framework of the RH problem with zeroes as described in the previous subsection.
Considering a trivial regular solution of the RH problem \eqref{eq:rh22} $J_0^\pm(x,t,k)=I_{n+1}$ which corresponds to  $Q(x,t)=0$,  
one then constructs  the corresponding RH problem with zeroes at $k^\pm_j$, 
$j=1,\dots,N$. The latter enjoys additional properties here due to the reduction symmetry (\ref{eq:laxpair1}). 
Precisely, \eqref{eq:det111} implies
\begin{eqnarray}
k_j^+=(k_j^-)^*\equiv k_j \in\CC^+\,,
\end{eqnarray}
One also gets 
\begin{eqnarray}
\label{eq:subspaces}
{\cal U}_j(x,t)=span\left\{e^{-i\phi(x,t,k^*_j) \Sigma_3}    \,  \bma  \beta_j  \\ -1 \ema\right\}~~,~~
{\cal V}^\perp_j(x,t)={\cal U}_j(x,t)\,,
\end{eqnarray}
where $\beta_j$ is a non-zero vector in $\CC^n$ and ${\cal V}^\perp_j$ represents the orthogonal complement of ${\cal V}_j$. Here $\beta_j$ 
is the so-called {\it norming constant} associated to $k_j$.
Note that \eqref{eq:subspaces} implies that the projectors involved in the dressing factors are 
rank-one orthogonal projectors. So the degree $1$ dressing factor reads 
\begin{equation}
  \label{eq:dm30}
  D_{i_j, \{i_1\dots i_{j-1}\}}(x,t,k) = 
  I_{n+1} + \left(f_{i_j}(k ) - 1\right) \Pi _{i_j, \{i_1\dots i_{j-1}\}}(x,t)\,, \quad f_{i_j}(k) = \frac{k-k_{i_j}}{ k-k^*_{i_j}}\,,
\end{equation}
and enjoys the property
\begin{equation}
  \label{eq:dm30}
  D_{i_j, \{i_1\dots i_{j-1}\}}^{-1}(x,t,k) = D_{i_j, \{i_1\dots i_{j-1}\}}^{\dagger}(x,t,k^*)\,.
\end{equation}
Defining
\begin{equation}
\label{eq:proj1122}
\zeta_{i_j,\{i_1\dots i_{j-1}\}}(x,t) = D^\dagger_{i_1\dots i_{j-1}}(x,t,k_{j})e^{-i\phi(x,t,k_{j}^* )\Sigma_3}\bma \beta_{i_j} \\ -1\ema\,,
\end{equation}
one gets
\begin{equation}
  \label{eq:proj111}
  \Pi_{i_j,\{i_1\dots i_{j-1}\}}(x,t) =  \frac{\zeta_{i_j,\{i_1\dots i_{j-1}\}} \zeta^\dagger_{i_j,\{i_1\dots i_{j-1}\}}(x,t)}
  {\zeta^\dagger_{i_j,\{i_1\dots i_{j-1}\}} \zeta_{i_j,\{i_1\dots i_{j-1}\}}(x,t)}\,. 
\end{equation}
In particular, the reconstruction formula \eqref{eq:rhh1} becomes 
\begin{equation}
  \label{eq:dm38}
  Q(x,t) = \sum_{j=1}^{N} i(k_{j} -k_{j}^*) [\Sigma_3, \Pi_{j, \{1,\dots, {j-1}\}}(x,t)]\,.
\end{equation}
For later convenience, we introduce the following definition.
\begin{definition}
\label{reduced_dressing}
Let $\sigma\in S_N$ be given and write $(\sigma(1),\dots,\sigma(N))=(i_1,\dots,i_N)$. Define $d_{i_1\dots i_\ell}(k)$, $1\le \ell\le N$, recursively 
by 
\begin{equation}
  \label{dm11}
    d_{i_1\dots i_\ell} (k) = d_{i_1}(k) \,d_{i_2, \{i_1\}}(k) \dots d_{i_\ell, \{i_1\dots i_{\ell-1}\}}(k) \,,
\end{equation}
where, for $1\le j\le \ell$,
\begin{equation}
  \label{eq:dm3000}
  d_{i_j, \{i_1\dots i_{j-1}\}}(k) = I_{n} + \left(f_{i_j}(k) - 1\right) \pi _{i_j, \{i_1\dots i_{j-1}\}}\,, 
\end{equation}
\begin{equation}
  \label{eq:proj1111}
  \pi_{i_j,\{i_1\dots i_{j-1}\}}=  \frac{\xi_{i_j, \{i_1\dots i_{j-1}\}} \xi^\dagger_{i_j, \{i_1\dots i_{j-1}\}}}{\xi^\dagger_{i_j, \{i_1\dots i_{j-1}\}} 
  \xi_{i_j, \{i_1\dots i_{j-1}\}}}\,, \quad \xi_{{i_j, \{i_1\dots i_{j-1}\}}} = d^\dagger_{\{i_1\dots i_{j-1}\}}(k_{i_j}) \beta_{i_j}\,, \quad f_l(k) = \frac{k - k_l}{k -  k^*_l}\, .
\end{equation}
\end{definition}
In the pure $N$-soliton system,  one has 
\begin{equation}
  \label{eq:detaa}
  a ^+ (k) =  d_{i_1 \dots i_N} (k)~~\text{and}~~\det a^+(k) =  \prod _{j=1}^N f_j(k) \,,
\end{equation}
where $a^+(k)$ is defined in \eqref{rep_S2}.
The analytic structure of $ a ^+ (k)$ is therefore completely determined by the dressing factor $ d_{i_1 \dots i_N} (k)$. Finally, we
introduce the matrix ${\cal A}_j$ by
\begin{equation}
  \displaystyle
  \label{eq:cst3}
 \frac{\mathcal{A}_j}{\det a^+(k_j)^{'}}  =  \lim_{k\to k_j} (k-k_j) (a^+(k))^{-1}  \,, \quad  \det a^+(k_j)^{'} = \left.\frac{d \det a^+(k) }{d k} \right|_{k= k _j} \,.
\end{equation}
${\cal A}_j$ contains the information on the residue of $(a^+(k))^{-1}$ at $k_j$ and plays an important role for the $N$-soliton solution of VNLS 
on the half-line \cite{2011arXiv1110.2990C}. It will be used in section \ref{sec:boundarycond}.

\subsection{Factorization of $ N$-soliton collision and Yang-Baxter maps}
\label{sec:facto}

We are now ready to discuss the factorization property of $N$-soliton collisions in the VNLS equation. 
This has already been treated in \cite{2004PThPh.111..151T,0266-5611-20-4-012} from two different angles. In \cite{2004PThPh.111..151T}, the 
factorization is shown by means of an involved direct computation from the $N$-soliton solution. 
In \cite{0266-5611-20-4-012}, two-soliton collisions are characterized by means of a Yang-Baxter map \cite{veselov2003yang}. 
Although the technical tool is based on asymptotic analyses of the $N$-soliton solution in both papers and also in ours, we stress that the 
difference here is that our approach is based on the dressing method and its important consequence Theorem \ref{prop23}. This ensures that the whole construction of the 
$N$-soliton solution is in fact consistent from the beginning. In a sense, the factorization of soliton collision is a consequence of this consistency. 
From this point of view, we have an \textit{a priori} proof of factorization whereas the discussions in \cite{2004PThPh.111..151T,0266-5611-20-4-012}
can be seen as \textit{a posteriori} checks that the $N$-soliton solution obtained with the ISM is consistent. 
The formulation in terms of Yang-Baxter maps turns out to be powerful to discuss factorization and provides a natural interpretation of
 dressing factors and their properties as we will see below. One important remark is that strictly speaking, in \cite{0266-5611-20-4-012}, 
the map between two polarization vectors is established from a two-soliton solution and then factorization is discussed based on this. 
However, to complete the argument, one has to derive such 
a map between two arbitrary polarization vectors within a full $N$-soliton solutions. This was in fact the main motivation for the alternative 
approach of \cite{2004PThPh.111..151T}. Here, we perform this task directly at the level of the dressing factors using Theorem \ref{prop23}. This 
results in particular in Lemma \ref{pro:122} below.

We now make the discussion more precise. First we need to introduce the general form of a one-soliton solution characterized by 
$k_0=\frac{1}{2}(u_0 + iv_0)$, $v_0>0$ and $\beta_0$,
\begin{eqnarray}
R(x,t)=\mathbf{p}_0\, v_0\frac{e^{-i(u_0x+(u_0^2-v_0^2)t)}}{\cosh(v_0 (x+2u_0 t-\Delta x_0))}\equiv \mathbf{p}_0 q_0(x,t)\,,
\end{eqnarray}
where $\Delta x_0=\frac{\ln |\beta_0|}{v_0}$, $\mathbf{p}_0=\frac{\beta_0}{|\beta_0|}$. The unit vector $\mathbf{p}_0$ 
is the \textit{polarization} of the soliton, $w_0=-2u_0$ its velocity, $v_0$ its amplitude and $\Delta x_0$ is the position of the maximum of the envelope of the soliton at $t=0$.
The main feature is that a vector one-soliton is simply a polarization vector $\mathbf{p}$ times a scalar one-soliton solution $q(x,t)$.

Now, consider the $N$-soliton solution corresponding to
\begin{equation}
  \label{eq:scatdata11}      
  \displaystyle
  k_j = \frac{1}{2}(u_j + iv_j)\, , \, v_j>0~~,~~j=1,\dots, N\,,
\end{equation}
with the associated norming constants $\beta_j$. Let $w_j = - 2 u_j$. Then, the following 
proposition shows that as $t\to\pm\infty$, an $N$-soliton solution looks like the sum of $N$ one-soliton solutions up to 
exponentially vanishing terms.
\begin{proposition}
\label{prop:30}
Suppose without loss of generality that $ u_1\, <\, u_2 < \, \dots\, <\,u_N$. 
Denote $R^{in}(x,t)$ (resp. $R^{out}(x,t)$) the asymptotic solution $R(x,t)$ corresponding to 
$t\to - \infty$ (resp. $t\to  \infty$).  Then, 
\begin{equation}
  \label{eq:asysol1}
  \displaystyle
  R^{in/out}(x,t) =\sum_{j=1}^N \mathbf{p}^{in/out}_j v_j\frac{e^{-i(u_jx+(u_j^2-v_j^2)t)}}{\cosh(v_j (x-w_j t-\Delta x_j^{in/out}))} + O (e^{-v \tilde{w}| t|} )\,.
\end{equation}
Here $v=\underset{j }{\min}\,v_j$, $\tilde{w} = \underset{l\neq j }{\min}\,  |w_l- w_j| $, $\Delta x_j^{in/out}=\frac{\ln |\beta_j^{in/out}|}{v_j}$ and $\mathbf{p}^{in/out}_j=\frac{\beta_j^{in/out}}{|\beta_j^{in/out}|}$ with
\begin{equation}
  \label{eq:fac63}
  \beta_{j}^{in}  =   \prod_{\ell=1}^{j-1} f_\ell(k^*_j)      \,  d^\dagger_{{j+1\dots N}} (k_j)\, \beta_{j}\,,      \quad
  \beta_j^{out}  =   \prod_{\ell=j+1}^N f_\ell(k^*_j)   \,  d^\dagger_{{1\dots j-1}}(k_j) \, \beta_{j}\,, 
\end{equation}
where $f_\ell(k)$ and $d_{{i_1\dots i_\ell}}(k)$ are defined in Sec. \ref{eq:reduction}.
\end{proposition}
\prf 
We follow the idea of the scalar case \cite{faddeev2007hamiltonian} which is based on the evaluation of the projectors $\Pi_{j, \{1\dots j-1\}}(x,t)$ as $t\to \pm \infty$. 
To get the result, it is enough 
to show that $R(x,t)$ approaches the one-soliton solution following the trajectory of a particular soliton $l$ i.e. $x- w_\ell t=constant$, and 
that it vanishes exponentially for all other directions in the $(x,t)$-plane. 
But here in the vector case, Theorem \ref{prop23} is crucial and allows us to write the dressing factor 
in the following form
\begin{equation}
  \label{eq:facto11}
  D_{1\dots N} =  D_1 \dots D_{\ell-1, \{1\dots \ell-2\}} D_{\ell+1, \{1 \dots \ell-1 \}} \dots D_{N, \{1\dots\hat{\ell}\dots N-1\}} D_{\ell, \{1\dots\hat{\ell}\dots N\}}\,,
\end{equation}
where the notation $\{1\dots\hat{\ell}\dots j\}$ means that $\ell$ is not listed in $\{1\dots j\}$.
This means that $D_{\ell, \{1\dots\hat{\ell}\dots N\}}$ is the last dressing factor added. Then, recalling (\ref{eq:proj1122}, \ref{eq:proj111}), 
one obtains for $x - w_\ell t=constant$ as $t \to - \infty$
\begin{equation}
  \label{eq:facto16}
 D_1  \dots 
   D_{N, \{1\dots\hat{\ell}\dots N-1\}}(x,t,k)= \bma d_{\ell+1\dots N}(k)  & 0 \\ 0 &  \prod_{j=1}^{\ell-1} f_j(k) \ema +  O (e^{-v \tilde{w}| t|} ) \,,
\end{equation}
whereas for all other directions, the same calculation yields $O (e^{-v \tilde{w}| t|} )$. Consequently, 
\begin{equation}
  \label{eq:facto17}
 \zeta_{\ell, \{1\dots\hat{\ell}\dots N\}}(x,t) =    \bma d^\dagger_{\ell+1\dots N}(k_\ell) & 0 \\ 0 &  \prod_{j=1}^{\ell-1} f^*_j(k_\ell) \ema   e^{- i \phi(x,t,k_\ell^*) \Sigma_3} 
   \bma\beta_\ell\\ -1\ema   + O (e^{-v \tilde{w}| t|} )\,,
\end{equation}
and the reconstruction formula \eqref{eq:dm38} implies
\begin{equation}
  \label{eq:facto171}
Q(x,t) = i(k_\ell - k_\ell^*)\left[ \Sigma_3, \frac{ \zeta_{\ell, \{1\dots\hat{\ell}\dots N\}}  \zeta_{\ell, \{1\dots\hat{\ell}\dots N\}} ^\dagger(x,t)}
{\zeta ^\dagger_{\ell, \{1\dots\hat{\ell}\dots N\}}   \zeta_{\ell, \{1\dots\hat{\ell}\dots N\}} (x,t)}\   \right]\,.
\end{equation}
Direct calculation then gives for $x - w_\ell t=constant$ as $t \to - \infty$
\begin{equation}
  \label{eq:facto172}
   R(x,t)= \mathbf{p}^{in}_\ell v_\ell\frac{e^{-i(u_\ell x+(u_\ell^2-v_\ell^2)t)}}{\cosh(v_\ell (x-w_\ell t-\Delta x_\ell^{in}))} + O (e^{-v \tilde{w}| t|} )\,,
\end{equation}
with the various parameters being defined in the proposition. The same technique can be applied as $t\to\infty$ to obtain $R^{out}(x,t)$.
\finprf

\begin{remark}
The order  $u_1<\dots<u_N$ means that the relative  velocity $w_j-w_{j+1}$ of two consecutive solitons is always positive.
Consequently, as $t \to - \infty$, the solitons are distributed along the x-axis in the order $1,2,\dots, N$. 
The picture is reversed as $t\to \infty$. The relative positions of the solitons are therefore completely determined as $t \to \pm \infty$.  
\end{remark}
\begin{remark}
Using Theorem \ref{prop23}, we could have performed the proof analogously but choosing any permutation placing $i_\ell$ at position 
$N$ hence giving 
$$D_{i_1} \dots D_{i_{\ell-1}, \{i_1\dots i_{\ell-2}\}} D_{i_\ell, \{i_1 \dots i_{\ell-1} \}} \dots D_{i_{N-1}, \{i_1\dots i_{N-2}\}} D_{i_\ell, \{i_1\dots i_{N-1}\}}$$ instead of 
\eqref{eq:facto11}. This corresponds to the possibility that the soliton collisions can occur in a different order since we do not know their relative positions at
an arbitrary time $t$.
However, the final result for $\beta_j^{in/out}$ would be the same. This is the essence of the 
factorization property. It turns out that this can be made precise by assigning an "intermediate time" polarization vector to each soliton and by 
considering the effect of a two-soliton collision {\bf within an $N$-soliton solution} on the assigned polarization vectors. 
The map between the polarization vectors before and after the two-soliton collision is a Yang-Baxter map satisfying the set-theoretical Yang-Baxter equation.
The mathematical translation of the factorization property of collisions is therefore an associativity property of the operation on polarization vectors given by the
Yang-Baxter map.
\end{remark}
\begin{remark}
The quantity $\Delta x^{out}_j-\Delta x^{in}_j$ represents the total position shift incurred by soliton $j$ through its collisions with the other solitons.  
The unit vector $\pp_j^{in/out}$ represents the asymptotic polarization vector of soliton $j$ before and after all its collisions with the other solitons. 
From the previous remark, these quantities are independent of the order of soliton collisions.
We see that by eliminating $\beta_j$ in \eqref{eq:fac63} that $\beta_j^{out}$ is completely determined by $\beta_j^{in}$ through the dressing factors. 
We note that the obtained relations are different from those obtained in the original paper by Manakov in that they do not call upon $\beta^{out}_j$ 
recursively for $j<\ell$ to obtain $\beta^{out}_\ell$. The original Manakov's formula made it extremely difficult to see the factorization property and in fact led him 
to conclude that it did not hold.
\end{remark}
To complete the argument and finish the proof of the claims in the previous remarks, we define the following "intermediate time" polarization vectors. Let
\begin{equation}
  \label{eq:fac62}
    \displaystyle
  \gamma_{i_j,\{ {i_{j+1}\dots i_N }\}}  =   \left(  \prod_{\ell=i_1}^{i_{j-1}} f_\ell(k^*_{i_j}) \right)  \,  d^\dagger_{i_{j+1}\dots i_N} (k_{i_j}) \, \beta_{i_	j}\,, \quad 
  \,,
\end{equation}
and
\begin{equation}
  \label{eq:facto18}
\pp_{i_j,\{ {i_{j+1}\dots i_N }\}}  = \frac{\gamma_{i_j,\{ {i_{j+1}\dots i_N }\}}  }{ |\gamma_{i_j,\{ {i_{j+1}\dots i_N }\}} |}\,.
\end{equation}
So in particular, $\pp^{in}_j=\pp_{j,\{ {j+1\dots N }\}}$ and $\pp^{out}_j=\pp_{j,\{ {1\dots j-1 }\}}$ and they can be pictorially represented as 
\begin{center}
  \begin{tikzpicture}[scale=0.8]
    \label{pic6}
    \draw[->][thick] (5,1) node[anchor=west]{$t \to -\infty$} -- (5,-1) node[anchor=west]  {$t \to \infty$}; 
    \node at (-3.1,0.8)[anchor=south east]{$\pp_{1,\{ 2\dots N \}}$};
    \node at (-1.3,0.8)[anchor=south  ]{$\pp_{j,\{ j+1 \dots N \}}$};
    \node at (3.1,0.8)[anchor=south west]{$\pp_{N}$};    
    \node at (1.4,0.8)[anchor=south ]{$\pp_{l,\{ l+1 \dots N \}}$};
    \node at (0,0.8)[anchor=south]{$..$}; 
    \node at (2.4,0.8)[anchor=south west]{$\dots$}; 
    \node at (-2.3,0.8)[anchor=south east]{$\dots$}; 
    \node at (0,-1)[anchor=north]{$..$}; 
    \node at (2.4,-1)[anchor=north west]{$\dots$}; 
    \node at (-2.3,-1)[anchor=north east]{$\dots$}; 
    \node at (-3.1,-0.8)[anchor=north east]{$\pp_{N,\{ 1 \dots N-1 \}}$};
    \node at (-1.3,-0.8)[anchor=north  ]{$\pp_{l,\{ 1 \dots l-1 \}}$};
    \node at (3.1,-0.8)[anchor=north west]{$\pp_{1}$};    
    \node at (1.4,-0.8)[anchor=north ]{$\pp_{j,\{ 1 \dots j-1 \}}$};
    \draw (0,0) node{{\small $N$-soliton collision}}ellipse (3cm and 0.6cm);
    \draw[->] (-1,-1.8) -- node[thick, anchor=north]{$x$} (1,-1.8);
  \end{tikzpicture}
\end{center}
We can now formulate the following important lemma.
\begin{lemma}
\label{pro:122}
Choose $k_{j}$ and $k_{l}$ and assume $u_{j}< u_{l}$. Write for convenience ${i}_\rho=i_1\dots i_q$ for some 
$q\in\{1,\dots,N\}$ such that $j$ and $l$ are not in $\{i_1,\dots,i_q\}$. Then
\begin{align}
  \label{eq:fac500}
  \mathbf{p}_{l,\{j \, {i}_\rho  \}} & = \frac{f^*_j(k^*_l)}{\Xi_{lj}} \left(  I_n + (f^*_j(k_l^*) - 1 ) \mathbf{p}_{j,\{l \, {i}_\rho\}} (\mathbf{p}_{j,\{l\, {i}_\rho\} })^\dagger \right)  
  \mathbf{p}_{l, \{ {i}_\rho \}} \,,\\
  \label{eq:fac501}
  \mathbf{p}_{j,\{{i}_\rho\}} & =  \frac{f_l(k^*_j)}{\Xi_{lj}} \left(  I_n + (f_l(k^*_j)-1 ) \mathbf{p}_{l,\{{i}_\rho\}} (\mathbf{p}_{l,\{{i}_\rho\}} )^\dagger \right)  
  \mathbf{p}_{j,\{l\, {i}_\rho\}} \,,
\end{align}
where
\begin{equation}
  \label{eq:fac53}
    \Xi_{lj}^2  =  \left|f_j(k_l^*) \right|^2 \left( 1 + \left(\frac{(k_j^*-k_j)(k_l -k_l^*)}{|k_l-k_j|^2} \right) | p_{jl,\{{i}_\rho\}}|^2 \right)\,, \quad  p_{jl,\{{i}_\rho\}} =  
    \mathbf{p}_{l,\{{i}_\rho\}}^\dagger  \mathbf{p}_{j,\{l\, {i}_\rho\}} \,.
\end{equation}    
\end{lemma}
\prf
From  \eqref{eq:fac62}, we have
\begin{align}
  \label{eq:fac531} 
  \gamma_{j, \{{i}_\rho\}}  =\underset{{p \in \{1\dots N\} \backslash \{j, {i}_\rho\}}}{ \prod f_p(k_j^*)} d^\dagger_{{i}_\rho}(k_j) \beta_j\,, \quad & 
  \gamma_{j,\{l \,{i}_\rho\}} = \underset{{p \in \{1\dots N \} \backslash \{j, l, {i}_\rho\}} }{\prod f_p(k_j^*) }d^\dagger_{l\,{i}_\rho}(k_j) \beta_j\,,\\
  \label{eq:fac533} 
  \gamma_{l, \{{i}_\rho\}}  = \underset{ {p \in \{1\dots N\} \backslash \{ l,{i}_\rho\}}}{\prod f_p(k_l^*) }d^\dagger_{{i}_\rho}(k_l) \beta_l\,, \quad  & 
  \gamma_{l,\{j\, {i}_\rho\}} = \underset{{p \in \{1\dots N \}\backslash \{l,j, {i}_\rho\}}}{\prod f_p(k_l^*)} d^\dagger_{j\,{i}_\rho}(k_l) \beta_l\,.
\end{align}
The relation between $d^\dagger_{{i}_\rho}$ and $d^\dagger_{j{i}_\rho}$ implies that 
\begin{align}
  \label{eq:fac535}
  \gamma_{j,\{l\, {i}_\rho\}} &= f^*_l(k_j) \left( I_n + (   f^*_l(k_j)  -1) \pp_{l,\{i_\rho\}} \pp^\dagger_{l,\{ i_\rho\}}  \right) \gamma_{j,\{{i}_\rho\}}\,,    \\
  \label{eq:fac536}
  \gamma_{l,\{j\, {i}_\rho\}} &= f^*_j(k_l) \left(  I_n + (
 f^*_j(k_l)  -1) \pp_{j,\{{i}_\rho\}} \pp^\dagger_{j,\{{i}_\rho\}} \right) \gamma_{l, \{{i}_\rho\}}\,.
\end{align}
Introduce 
\begin{equation}
  \label{eq:fac537}
  \Xi_{lj} = \frac{|\gamma_{j, \{{i}_\rho\}}|}{| \gamma_{j, \{l\, {i}_\rho\}}|}\,,
\end{equation}
we have
\begin{align}
  \displaystyle
  \label{eq:fac538}
  \Xi_{lj}^2 
    &=\frac{|f_l(k^*_j)|^2}{|\gamma_{j,\{l\,{i}_\rho\}}|^2}\left( \gamma^\dagger_{j,\{l {i}_\rho\}}\left(I_n + f_l^*(k^*_j)\pp_{l,\{{i}_\rho\}} 
    \pp^\dagger_{l,\{{i}_\rho\}}\right)   
      \left(I_n + f_l(k^*_j)\pp_{l,\{{i}_\rho\}} \pp^\dagger_{l,\{{i}_\rho\}}\right)     \gamma_{j,\{l\, {i}_\rho\}} \right)\,, \nonumber\\
  &=|f_l(k^*_j)|^2\left(1 + \frac{(k_l - k^*_l)(k_j^*-k_j)}{|k_l - k_j|^2} |p_{jl,\{{i}_\rho\}} |^2 \right)\,.
\end{align}
It is easy to see that $\Xi_{lj} = \Xi_{jl}$. Inserting \eqref{eq:fac537} into (\ref{eq:fac535}, \ref{eq:fac536}) yields (\ref{eq:fac500}, \ref{eq:fac501}) by direct calculation.
\finprf
\begin{remark}
The relations defined in Lemma \ref{pro:122} have a natural interpretation as an "intermediate time" pairwise collision between soliton $j$ and soliton $l$. 
Since $w_j > w_l$ ($u_j < u_l$), after  a certain number of collisions with other solitons (related to the set $\{{i}_\rho\}$), soliton $j$ with polarization $\pp_{j, \{l i_\rho\}}$ 
overtakes soliton $l$ with polarization $\pp_{l, \{{i}_\rho\}}$ and acquires polarization $\pp_{j, \{{i}_\rho\}}$ while soliton $l$ has then polarization $\pp_{l, \{j {i}_\rho\}}$. 
Pictorially, this can be represented as
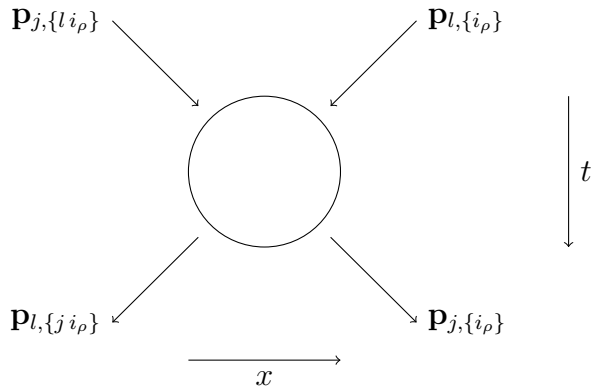
\begin{center}
  \begin{tikzpicture}
    \draw[<-] (-0.87,0.87) --   (-2,2) node[anchor= east] {$\pp_{j, \{l\,i_\rho \}}$} ;
    \draw[->] (-0.87,-0.87) --  (-2,-2) node[anchor= east] {$\pp_{l, \{j\, i_\rho \}}$};
    \draw[<-] (0.87,0.87) --   (2,2) node[anchor=west] {$\pp_{l, \{ i_\rho  \}}$}; 
    \draw[->] (0.87,-0.87) --  (2,-2) node[anchor=west] {$\pp_{j, \{ i_\rho \}}$} ;
    \draw (0,0)  circle (1cm) ;
    \draw[->] (-1,-2.5) -- node[thick, anchor = north]{$x$} (1,-2.5);
    \draw[->] (4,1) -- node[thick, anchor = west]{$t$} (4,-1);
  \end{tikzpicture}
\end{center}
\end{remark}
We complete the discussion by rewriting the relations in Lemma \ref{pro:122} in terms of the following map acting on $\CC\PP^{n-1}\times \CC\PP^{n-1}$
onto itself (so that the normalizations in \eqref{eq:fac500}, \eqref{eq:fac501} are irrelevant)
\begin{equation}
  \label{eq:yba201}
                 {\cal R}(k_1, k_2 ) : (\pp^{(\romannumeral 1)}_1 , \pp^{(\romannumeral 1)}_2 ) \mapsto  (\pp^{(\romannumeral 2)}_1 , \pp^{(\romannumeral 2)}_2 )\,,
\end{equation}
\begin{align}
  \label{eq:fac690}
  \mathbf{p}^{(\romannumeral 2)} _{1}    =  \left(  I_n +  \left( \frac{k_1^* - k_2}{k_1^* - k_2^*}  - 1\right) 
  \frac{\pp^{(\romannumeral 1)}_2 (\pp^{(\romannumeral 1)}_2)^\dagger}{(\pp^{(\romannumeral 1)}_2)^\dagger \pp^{(\romannumeral 1)}_2}      \right) \pp^{(\romannumeral 1)}_1 \,, \\
  \label{eq:fac691}
  \mathbf{p}^{(\romannumeral 2)} _{2}    =  \left(  I_n +  \left( \frac{k_2 - k_1^*}{k_2 - k_1}  - 1\right) 
  \frac{\pp^{(\romannumeral 1)}_1 (\pp^{(\romannumeral 1)}_1)^\dagger}{(\pp^{(\romannumeral 1)}_1)^\dagger \pp^{(\romannumeral 1)}_1}      \right) \pp^{(\romannumeral 1)}_2 \,.
\end{align}
Then, one finds that the map ${\cal R}(k_1,k_2)$ is a reversible (parametric) Yang-Baxter map \cite{veselov2003yang} i.e. it satisfies
\footnote{This is introduced properly in section \ref{RM}.}
\begin{equation}
  \label{eq:yba114}
   {\cal R}_{12}(k_1, k_2)   {\cal R}_{13}(k_1, k_3)   {\cal R}_{23}(k_2,k_3) = {\cal R}_{23} (k_2,k_3)   {\cal R}_{13} (k_1,k_3)  {\cal R}_{12}(k_1,k_2)\,,
\end{equation}
and 
\begin{equation}
  \label{eq:yba115}
   {\cal R}_{21}(k_2, k_1)   {\cal R} (k_1, k_2) = Id\,.
\end{equation}
The rewriting of relations between polarization vectors as a Yang-Baxter map is the basis of the argument in \cite{0266-5611-20-4-012}. 
We stress however that the crucial difference of our approach is that we obtained the Yang-Baxter map in complete generality 
for arbitrary polarization vectors within a full $N$-soliton solution and not just from the two-soliton solution. This is similar in spirit to the approach by Tsuchida in 
\cite{2004PThPh.111..151T} but again with the importance difference that here, this was made possible by our {\bf a priori} derivation of Theorem \ref{prop23} about dressing factors, 
instead of an {\bf a posteriori} derivation from the explicit $N$-soliton solution. We then recover the following result originally formulated in \cite{2004PThPh.111..151T}
\begin{theorem}
\label{theo11}
An $N$-soliton collision in the Manakov model can be factorized into a nonlinear superposition of $\bma N \\ 2\ema$ pairwise collisions in  arbitary order.
\end{theorem}

\section{Factorization with an integrable boundary}
\label{sec:boundarycond}

In this section, we use the main results of \cite{2011arXiv1110.2990C} and combine them with the previous construction to discuss factorization of 
 $N$-soliton solutions of VNLS on the half-line. The idea is that such a solution can be obtained as the restriction to $x>0$ of a $2N$-soliton solution of VNLS 
 on the full line provided that the norming constants and the poles $k_j$ obey suitable mirror symmetry conditions depending on the boundary conditions 
 \eqref{Robin_BC} or \eqref{mixed_BC1}, \eqref{mixed_BC2}. 

 To fix ideas, consider $k_j=\frac{1}{2}(u_j+iv_j)$, $v_j>0$, $j=1,\dots,2N$ and assume that 
\begin{align}
  \label{eq:a23}
  u_j >0\, , \text{ for } j =  1, \dots , N\quad \text{and} \quad & u_1 < u_2 < \dots < u_N\,.
  \end{align}
This corresponds to the situation where solitons $1$ to $N$ are the "real" solitons (on $x> 0$) and the solitons $N+1$ to $2N$ are the "mirror" solitons (on $x<0$) as $t\to-\infty$. 
The "real" solitons have negative velocities so they evolve towards the boundary where they meet their "mirror"  solitons which then become the real solitons. The net result when restricted 
to $x>0$ is that $N$ solitons interact with the boundary and bounce back. This looks like 
\begin{align*}
  2N\,,\, 2N-1\,,\,  \dots \,, N+1\, \Big|  \, 1\,, \, 2\,, \,\dots \, N\,,   \quad t\to -\infty\,, \\
  N\,,\, N-1\,,\,  \dots \,, 1\, \Big|  \, N+ 1\,, \, N+2\,, \,\dots \, 2N\,,   \quad t\to \infty\,,
\end{align*}
where the vertical bar represents the boundary. 
We can now state the results from \cite{2011arXiv1110.2990C} in a form convenient for our purposes here.
\begin{proposition}
The $N$-soliton solution of VNLS on the half-line with boundary conditions \eqref{Robin_BC} or \eqref{mixed_BC1}, \eqref{mixed_BC2} is obtained by the dressing 
procedure described in Sec. \ref{eq:reduction} based on $k_j$ and $\beta_j$, $j=1,\dots,2N$ with $k_j$ given by \eqref{eq:a23} for $j=1,\dots, N$ and the following constraints
\begin{equation}
  \label{eq:aaa24}
  k_{j+N} =  - k_j^* \,, \quad  \beta_j \beta^\dagger_{j+N} = M(k_j^*) {\cal A}_{j+N}\,, \quad  \text{ for } j= 1, \dots , N\,,
\end{equation}
where 
\begin{align}
   \label{eq:pp12}
M(k) = \frac{k-i\alpha}{k+i\alpha} I_n\,, \quad \alpha \in \RR\,, \\
\text{or} \qquad \qquad \qquad \qquad \qquad \qquad \qquad \qquad & \nonumber \\
   \label{eq:pp13}
 M(k) = \bma \sigma_1  &  &  \\ & \ddots   & \\ &  &  \sigma_n   \ema\,, \quad   \begin{aligned} & \sigma_j = 1\,, j \in S\,,  \\ & \sigma_l =-1\,,  l \in \{1\dots n\}\backslash S   \,,
\end{aligned} 
\end{align}
and ${\cal A}_{j+N}$ is defined in \eqref{eq:cst3}.
\end{proposition}
The crux of the matter is to solve for $\beta_{j+N}$ given $\beta_j$ using \eqref{eq:aaa24}. Indeed, the definition of ${\cal A}_{j+N}$ leads to the  following expression
\begin{align}
  {\cal A}_{j+N}=  \prod_{i =1, i \neq j+N}^{2N}\left(  \frac{k_{j+N} - k_i}{k_{j+N} - k_i^*} \right) &  \left( d^{\dagger}_{2N, \{1 \dots 2N-1\}}  \, \dots \, d^{\dagger}_{j+N+1, \{1 \dots j+N\}} 
  \pi_{j+N, \{1\dots j+N-1\}} \right. \nonumber
\\   \label{eq:rela41} 
&\left.   \, d^{\dagger}_{j+N-1, \{1\dots j+N-2\}} \, \dots \, d^{\dagger}_{1} \right) (k_{j+N}) \,,
\end{align}
where $d_{i_j,\{i_1\dots i_{j-1}\}}(k)$ was introduced in Def. \ref{reduced_dressing}. Hence, the equations in \eqref{eq:aaa24} are coupled nonlinear equations for the $\beta_j$'s. 
In Appendix \ref{algo}, we provide an algorithm to solve them. With the knowledge of all 
the norming constants, we have access to the "intermediate-time" polarization vectors which are the basis for the proof of factorization on the half-line. They enjoy the following 
property.
\begin{proposition}
\label{prop:refmap1}
Consider $2N$ polarization vectors as defined in (\ref{eq:fac62}, \ref{eq:facto18}). 
Let $\{i_1 \dots i_N\}$ be a permutation of $\{ 1 \dots N\}$ and $\{k_{i_j}, k_{i_{j}+N};  \beta_{i_j}, \beta_{i_{j}+N}\}_{j \in \{1\dots N\}} $ be the corresponding 
"real" and "mirror" poles and norming constants satisfying
\begin{equation}
  \label{eq:a24}
  k_{i_{j}+N} =  - k_{i_j}^* \,, \quad  \beta_{i_j} \beta^\dagger_{i_{j}+N} = M(k_{i_j}^*) {\cal A}_{i_{j}+N}\,.
\end{equation}
Then the following relations hold 
\begin{equation}
  \label{eq:d11}
    \pp_{i_{j}+N, \{i_1\dots i_N\,i_{1}+N\dots i_{j-1}+N\}} =  \mm(k_{i_j})\,   \pp_ {i_j,  \{i_{j+1}\dots i_N\}}\,, 
\end{equation}
with the $n\times n$ matrix function   $\mathbf{m}(k)  $  defined by
\begin{align}
  \label{eq:ref2222}
  \mathbf{m}(k)  =  \frac{h(k)}{|h(k)|} I_{n}\,, \quad h(k) = \left(\frac{k - i\alpha  }{k+i\alpha }\right)\,, \\
 \text{or}  \qquad \qquad \qquad \qquad \qquad \qquad  \qquad \qquad \qquad & \nonumber \\
  \label{eq:ref22222}
\mathbf{m}(k) =  \bma \sigma_1 & & \\ & \ddots & \\ & & \sigma_n \ema\,, \quad   \begin{aligned} & \sigma_{p} = 1\,, p \in S\,,  \\ & \sigma_{q} =-1\,,  q \in \{1\dots n\}\backslash S  \,.
\end{aligned}  
\end{align}
In particular,
\begin{align}
  \label{eq:c11}
  \pp_{{j+N}, \{{1\dots N}\} }\,   =   \, \mathbf{m}(k_{j}) \, \pp_{j, \{1\dots\hat{j}\dots N\}}\,. 
\end{align}
\end{proposition}
The proof is long and is given in Appendix \ref{proof_prop_mirror}.
Let us look at these relations from the soliton collision viewpoint. As $t\to -\infty$, the polarization vectors are ordered as follows along the $x$-axis 
\[
  \label{eq:a31}
  \pp_{2N, \{1\dots 2N-1\}}\,,\, \pp_{2N-1, \{ 1\dots 2N-2\}}\,,\,  \dots \,, \pp_{N+1, \{1\dots N\}}\, \Big|\, \pp_{1,\{2 \dots N \}}\,, \, \pp_{2,\{ 3\dots N\}}\,, \,\dots \, \pp_{N}\,, \quad t\to -\infty\,.
\]
By applying \eqref{eq:d11}, this becomes
\[
  \label{eq:a31}
  \mm(k_{N}) \, \pp_{N}\,,\,   \dots \,,\mm(k_{2})\,  \pp_{2, \{3 \dots N\}}\, , \mm(k_{1}) \,\pp_{1, \{2\dots N\}}\, \Big|\,  \pp_{1,\{2 \dots N \}}\,, \, \pp_{2,\{ 3\dots N\}}\,, \,\dots \, \pp_{N}\,, \quad t\to-\infty\,.
\]
In fact, \eqref{eq:d11} shows that this picture extends to all "intermediate-time" polarization vectors. Therefore, any pairwise collision between soliton 
$i_l$ and $i_j$,   described by the Yang-Baxter map $\yR_{i_li_j}(k_{i_l}, k_{i_j})$, is accompanied by a "simultaneous" pairwise collision between solitons 
$i_{j}+N$ and $i_{l}+N$ described by $\yR_{i_{j}+N\, i_{l}+N}(k_{i_{j}+N}, k_{i_{l}+N})$, and vice versa.

Consider now the situation evolving from $t\to-\infty$. After a certain number of pairwise collisions, soliton $j$ is next to the boundary and the pairwise collision that takes place 
is given by $\yR_{j+N\, j}(k_{j+N}, k_j)$. The map $\yR_{j+N\, j}(k_{j+N}, k_j)$ is naturally interpreted as the reflection map of soliton $j$ on the boundary.
After the reflection, soliton $j+N$, now playing the role of the reflected soliton $j$, undergoes general pairwise collisions of the form 
$\yR_{j+N\, l}(k_{j+N}, k_l)$ with the remaining "real" solitons and $\yR_{q+N\, j+N}(k_{q+N}, k_{j+N})$  with the "mirror" solitons that travel faster than it.

We can now define the following \textit{reflection map} from $\CC\PP^{n-1}\times (\CC\setminus i\RR)$ to itself that describes the change of the polarization vector of one soliton when 
it interacts with the boundary 
\begin{equation}
  \label{eq:d13}
                 {\rR} : (\pp, k ) \mapsto   (\rp, -k^*)\,,
\end{equation}
where 
 \begin{equation}
   \label{eq:ref11}
   \rp  = \left(  I_n + \frac{k-k^*}{k+k^*} \frac{\pp \pp^\dagger}{\pp^\dagger \pp} \right) \,\mathbf{m}(k) \,\pp \,.
 \end{equation}
Pictorially, this corresponds to 
\begin{center}
  \begin{tikzpicture}[scale=1.3]
    \draw [dashed] (-2, 1.5) -- (-2,-1.5);
    \draw [thick] (-1,1) node[anchor=south west]{$\pp_{j, \{1\dots\hat{j}\dots N\}} $} -- (-2,0) node[anchor = west]{  \quad $ \yR_{j+N\,  j}(k_{j+N},k_j)$} -- (-1,-1) 
    node[anchor=north west]{$\pp_{j+N, \{1\dots\hat{j}\dots N\} }$ } ;
    \draw  [thick] (-3,1) node[anchor=south east]{$\pp_{j+N, \{ 1\dots N\} } $} -- (-2,0)  -- (-3,-1) node[anchor=north east]{$\pp_{j, \{j+N1\dots\hat{j}\dots N\} }$ } ;
    \node at (1.2,0)[thick] {$\to$};
  \draw [thick] (2, 1.5) -- (2,-1.5);
    \draw [thick] (3,1) node[anchor=south west]{$\pp_j $} -- (2,0) node[anchor = west]{  \quad $ \rR_{j}(k_j)$} -- (3,-1) node[anchor=north west]{$\rp_j$ } ;
  
  \end{tikzpicture}
\end{center}
For convenience, we introduce a (generalized) parametric notation\footnote{This is defined properly in section \ref{RM}.} for the reflection map: $\rR(k)$ acting only on vector $\pp$. We also define its action on a multiplet of vectors
\begin{equation}
\label{boundary_j}
  {\rR}_j(k_j) : (\pp_1,\dots,\pp_j,\dots,\pp_N) \mapsto (\pp_1,\dots,\rp_j,\dots,\pp_N)\,.
\end{equation}
We can now state the main theorem of this paper which guarantees the factorization property of the soliton collisions with a boundary.
\begin{theorem}
\label{prop:refmap2}
Consider the Yang-Baxter map $\yR_{lj}(k_l, k_j)$ defined in (\ref{eq:yba201} - \ref{eq:fac691}) and $\rR_j(k_j)$ in (\ref{eq:d13} - \ref{boundary_j}). Then the following set theoretical 
(parametric) reflection equation holds as an identity of maps on $\CC\PP^{n-1}\times \CC\PP^{n-1}$
\begin{align}
  \label{eq:ref14}
  \rR _1 (k_1 )   \yR_{21} ( -k_2^*, k_1)  & \rR_2 (k_2)   {\cal R}_{12}(k_1,k_2)   =\nonumber \\ &\yR_{21} (-k_2^*, -k_1^*)    \rR_2 (k_2)  \yR _{12} (-k_1^*,k_2)   \rR_1 (k_1 )\,.
\end{align}
Moreover, the parametric reflection map $\rR(k)$ satisfies the following involutive property
\begin{equation}
\label{eq:f11}
 \rR(- k^*)  \rR(k) = Id\,.
\end{equation}
\end{theorem}
\prf The involutive property can be found by direct calculation from the definition of the reflection map. 
For the set-theoretical reflection equation, we use the set theoretical Yang-Baxter equation together with our mirror image picture. 
Take $k_j$ and $k_{j+2} = -k_j^*$, $j=1,2$. Since pairwise collision occurs simultaneously on each side of the boundary, 
there are only $2$ possible configurations of collisions and they are identical using the Yang-Baxter equation
\begin{align}
\label{4YB}
\yR_{31}(k_1,k_3) & \yR_{32} (k_3,k_2) \yR_{41} (k_4,k_1) \yR_{42}(k_4,k_2) \yR_{43}(k_4,k_3)  \yR_{12}(k_1,k_2) = \nonumber  \\  
& \yR_{43} (k_4,k_3) \yR_{12}(k_1,k_2)  \yR_{42}(k_4,k_2)  \yR_{32}(k_4,k_2)  \yR_{41}(k_4,k_1)  \yR_{31} (k_3,k_1) \,.
\end{align}
Consider now the following sequence of pairwise collisions
\begin{align}
\label{eq:f131}
 &\yR_{31}(k_3,k_1)  \yR_{32} (k_3,k_2) \yR_{41} (k_4,k_1) \yR_{42}(k_4,k_2)  \yR_{43}(k_4,k_3)  \yR_{12}(k_1,k_2) (\pp^{(\romannumeral 1)}_{1},\pp^{(\romannumeral 1)}_{2},\pp^{(\romannumeral 1)}_{3}, \pp^{(\romannumeral 1)}_{4}  ) \nonumber \\
 = &\yR_{31}(k_3,k_1)  \yR_{32} (k_3,k_2) \yR_{41} (k_4,k_1) \yR_{42}(k_4,k_2)     (\pp^{(\romannumeral 2)}_{1},\pp^{(\romannumeral 2)}_{2},\pp^{(\romannumeral 2)}_{3}, \pp^{(\romannumeral 2)}_{4}  ) \nonumber  \\
 = &\yR_{31}(k_3,k_1)  \yR_{32} (k_3,k_2) \yR_{41} (k_4,k_1)      (\pp^{(\romannumeral 3)}_{1},\pp^{(\romannumeral 3)}_{2},\pp^{(\romannumeral 3)}_{3}, \pp^{(\romannumeral 3)}_{4}  ) \nonumber  \\
=& \yR_{31}(k_3,k_1)  (\pp^{(\romannumeral 4)}_{1},\pp^{(\romannumeral 4)}_{2},\pp^{(\romannumeral 4)}_{3}, \pp^{(\romannumeral 4)}_{4}  )\nonumber  \\
=& (\pp^{(\romannumeral 5)}_{1},\pp^{(\romannumeral 5)}_{2},\pp^{(\romannumeral 5)}_{3}, \pp^{(\romannumeral 5)}_{4}  )\,,
\end{align}
with $(\pp^{(\romannumeral 1)}_{1},\pp^{(\romannumeral 1)}_{2},\pp^{(\romannumeral 1)}_{3}, \pp^{(\romannumeral 1)}_{4}  ) $ being the initial polarization vectors and 
$(\pp^{(\romannumeral 5)}_{1},\pp^{(\romannumeral 5)}_{2},\pp^{(\romannumeral 5)}_{3}, \pp^{(\romannumeral 5)}_{4}  ) $ the final polarization vectors. 
Similarly, consider the following sequence of soliton-soliton and soliton-boundary collisions
\begin{align}
\label{eq:f132}
 & \rR _1 (k_1 )   \yR_{21} ( -k_2^*, k_1)  \rR_2 (k_2)   {\cal R}_{12}(k_1,k_2) ( \qq^{(\romannumeral 1)}_1, \qq^{(\romannumeral 1)}_2)\nonumber \\
= & \rR _1 (k_1 )   \yR_{21} ( -k_2^*, k_1)  \rR_2 (k_2)   ( \qq^{(\romannumeral 2)}_1, \qq^{(\romannumeral 2)}_2)\nonumber \\
=& \rR _1 (k_1 )   \yR_{21} ( -k_2^*, k_1)      ( \qq^{(\romannumeral 3)}_1, \qq^{(\romannumeral 3)}_2)\nonumber \\
=& \rR _1 (k_1 )    ( \qq^{(\romannumeral 4)}_1, \qq^{(\romannumeral 4)}_2)\nonumber \\
=&  ( \qq^{(\romannumeral 5)}_1, \qq^{(\romannumeral 5)}_2)\,.
\end{align}
We claim now that if $( \qq^{(\romannumeral 1)}_1, \qq^{(\romannumeral 1)}_2) =( \pp^{(\romannumeral 1)}_1, \pp^{(\romannumeral 1)}_2)$, 
then $( \qq^{(\romannumeral 5)}_1, \qq^{(\romannumeral 5)}_2) =( \pp^{(\romannumeral 5)}_3, \pp^{(\romannumeral 5)}_4)$. 
From \eqref{eq:f131} and  \eqref{eq:f132}, we have 
\begin{align}
  \label{reflec11}
  \pp^{(\romannumeral 2)}_1 & = \left(  I_n +  \left (\frac{k_2^* - k_2}{k_1^* - k_2^*}  \right) 
  \frac{\pp_2^{(\romannumeral 1)} (\pp^{(\romannumeral 1)}_2)^\dagger}{(\pp^{(\romannumeral 1)}_2)^\dagger\pp_2^{(\romannumeral 1)} } \right)   \pp_1^{(\romannumeral 1)}\,, \\  
  \label{reflec12}
  \qq^{(\romannumeral 2)}_1 & = \left(  I_n +  \left( \frac{k^*_2 - k_2}{k_1^* - k_2^*}  \right) 
  \frac{\qq_2^{(\romannumeral 1)} (\qq^{(\romannumeral 1)}_2)^\dagger}{(\qq^{(\romannumeral 1)}_2)^\dagger\qq_2^{(\romannumeral 1)} } \right)   \qq_1^{(\romannumeral 1)}\,, \\  
  \label{reflec13}
  \pp^{(\romannumeral 4)}_1 & = \left(  I_n +  \left( \frac{k_4 - k_4^*}{k_1 - k_4}   \right) 
  \frac{\pp^{(\romannumeral 3)}_4 (\pp^{(\romannumeral 3)}_4)^\dagger}{(\pp^{(\romannumeral 3)}_4)^\dagger\pp^{(\romannumeral 3)}_4 } \right) \pp^{(\romannumeral 3)}_1 \,, \\
  \label{reflec14}
  \qq^{(\romannumeral 4)}_1 & = \left(  I_n +  \left( \frac{k_2 - k_2^*}{k_1 + k_2^*}  \right) 
  \frac{\qq^{(\romannumeral 3)}_2 (\qq^{(\romannumeral 3)}_2)^\dagger}{(\qq^{(\romannumeral 3)}_2)^\dagger\qq^{(\romannumeral 3)}_2 } \right) \qq^{(\romannumeral 3)}_1 \,, \\
  \label{reflec15}
  \pp^{(\romannumeral 5)}_3 & = \left(  I_n +  \left( \frac{k_1^* - k_1}{k_3^* - k_1^*}  \right) 
  \frac{\pp^{(\romannumeral 4)}_1 (\pp^{(\romannumeral 4)}_1)^\dagger}{(\pp^{(\romannumeral 4)}_1)^\dagger\pp^{(\romannumeral 4)}_1 } \right) \pp^{(\romannumeral 4)}_3 \,, \\
  \label{reflec16}
  \qq^{(\romannumeral 5)}_1 & = \left(  I_n +  \left( \frac{k_1 - k_1^*}{k_1 + k_1^*}  \right) 
  \frac{\qq^{(\romannumeral 4)}_1 (\qq^{(\romannumeral 4)}_1)^\dagger}{(\qq^{(\romannumeral 4)}_1)^\dagger\qq^{(\romannumeral 4)}_1 } \right) \mm(k_1)\qq^{(\romannumeral 4)}_1 \,.
\end{align}
First, from  (\ref{reflec11}, \ref{reflec12}) we have  $ \pp^{(\romannumeral 2)}_1 =  \qq^{(\romannumeral 2)}_1$, since $\pp^{(\romannumeral 1)}_j = \qq^{(\romannumeral 1)}_j$, $j = 1,2$. 
Then, we have $\pp^{(\romannumeral 3)}_1 =  \pp^{(\romannumeral 2)}_1$ and $\qq^{(\romannumeral 3)}_1 =  \qq^{(\romannumeral 2)}_1 $. 
Next, recall that $k_4 = -k_2^*$ and  $\pp^{(\romannumeral 2)}_4 = \mm(k_2) \pp^{(\romannumeral 2)}_2$ so $\pp^{(\romannumeral 3)}_4 =  \qq^{(\romannumeral 3)}_2$.  
(\ref{reflec13}, \ref{reflec14}) give $\pp^{(\romannumeral 4)}_1 =  \qq^{(\romannumeral 4)}_1$.  Finally, since $k_3 = -k^*_1$ and $\pp^{(\romannumeral 4)}_3 = \mm(k_1) \pp^{(\romannumeral 4)}_1$, 
relations (\ref{reflec15}, \ref{reflec16})  imply  $ \pp^{(\romannumeral 5)}_3 =  \qq^{(\romannumeral 5)}_1$. It is easy to check  $ \pp^{(\romannumeral 5)}_4 =  \qq^{(\romannumeral 5)}_2$ as well. 
The same argument holds for
\begin{align}
\label{eq:f133}
&\yR_{12}(k_1,k_2)  \yR_{43} (k_4,k_3) \yR_{42} (k_4,k_2) \yR_{32}(k_3,k_2) \yR_{41}(k_4,k_1)  \yR_{13}(k_1,k_3)(\pp^{(\romannumeral 1)}_{1},\pp^{(\romannumeral 1)}_{2},
\pp^{(\romannumeral 1)}_{3}, \pp^{(\romannumeral 1)}_{4}   )\,,
\end{align} 
and
\begin{align}
\label{eq:f134}
&\yR_{21} (-k^*_2,-k^*_1) \rR_{2}(k_2) \yR_{12}(-k^*_1,k_2)  \rR_{1}(k_1)    (\qq^{(\romannumeral 1)}_{1},\qq^{(\romannumeral 1)}_{2}  ) \,.
\end{align} 
Since  \eqref{eq:f131} is equal to \eqref{eq:f133} by \eqref{4YB}, then  \eqref{eq:f132} is equal to \eqref{eq:f134}. \eqref{eq:f11} can be showed by the same way by considering the unitary property of $R_{12}(k_1,k_2)$ defined in \eqref{eq:yba115} and the proof is complete.
\finprf
The previous proof amounts to the following diagram
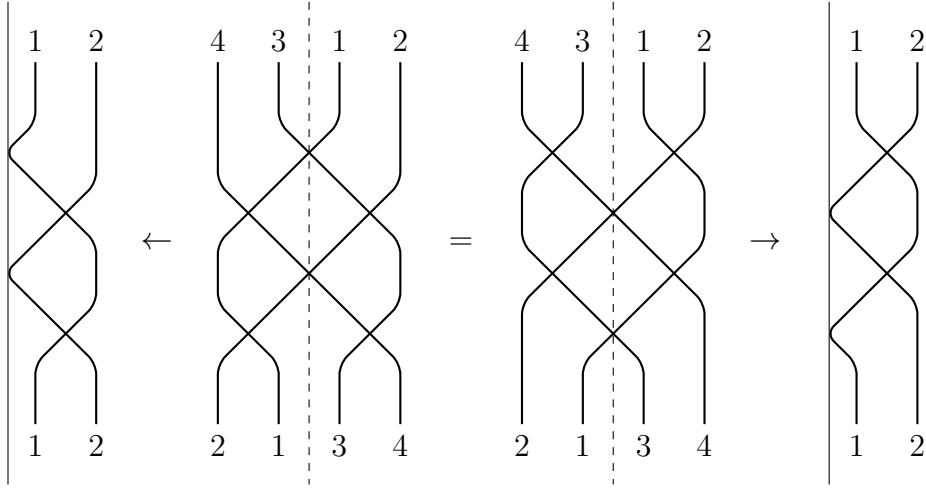
\begin{center}
  \begin{tikzpicture}[scale=0.8]
    \draw  (1.05,4) -- (1.05,-4);
    \draw [thick,rounded corners=4pt] (-4,3) node[anchor=south]{$4 $} -- (-4,2) --  (-1,-1)  -- (-1,-3) node[anchor=north]{$4$   } ;
    \draw  [thick,rounded corners=4pt] (-3,3) node[anchor=south ]{$3$ } -- (-3,2) -- (-4,1)  -- (-4,0)  -- (-2,-2) -- (-2,-3) node[anchor=north]{$3 $   } ;
    \draw [thick,rounded corners=4pt] (-2,3) node[anchor=south ]{$1 $} -- (-2,2)  -- (-1,1)  -- (-1,0) -- (-3,-2) -- (-3, -3) node[anchor=north ]{$1 $} ;
    \draw [thick,rounded corners=4pt] (-1,3) node[anchor=south ]{$2 $}  -- (-1,2) -- (-4,-1) -- (-4,-3)node[anchor=north ]{$2 $}  ;
    
   \node at (0,0)[thick] {$\to$};
    \node at (-10,0)[thick] {$\gets$};
    \node at (-5,0)[thick] {$=$};

   \draw [thick,rounded corners=4pt] (1.5,3)  node[anchor=south]{$1 $   }   -- (1.5,2)  -- (2.5,1)  -- (2.5,0) -- (1,-1.5) -- (1.5, -2) -- (1.5,-3)node[anchor=north ]{$1 $}  ;
   \draw [thick,rounded corners=4pt] (2.5,3)  node[anchor=south]{$2 $   }    -- (2.5,2)  -- (1,0.5) -- (2.5,-1)  -- (2.5,-3)node[anchor=north ]{$2 $}  ;
   

    \draw [thick,rounded corners=4pt] (-9,3) node[anchor=south]{$4 $   } -- (-9,1) --  (-6,-2)  -- (-6,-3) node[anchor=north]{$4 $   } ;
    \draw  [thick,rounded corners=4pt] (-8,3) node[anchor=south]{$3 $   }  -- (-8,2) -- (-6,0)  -- (-6,-1) -- (-7,-2)  -- (-7,-3) node[anchor=north]{$3 $   } ;
    \draw [thick,rounded corners=4pt] (-7,3)  node[anchor=south]{$1 $   }  -- (-7,2)  -- (-9,-0)  -- (-9,-1) -- (-8,-2) -- (-8, -3) node[anchor=north]{$1$   } ;
   \draw [thick,rounded corners=4pt] (-6,3)  node[anchor=south]{$2 $   }  -- (-6,1) -- (-9,-2) -- (-9,-3)  node[anchor=north]{$2 $   } ;
\draw [dashed] (-7.5,4) -- (-7.5,-4);
\draw [dashed] (-2.5,4) -- (-2.5,-4);
   \draw (-12.45,4) -- (-12.45,-4);
%
   \draw [thick,rounded corners=4pt] (-12,3)  node[anchor=south]{$1 $   } --  (-12,2)-- (-12.5,1.5)  -- (-11,0)  -- (-11,-1) -- (-12, -2) -- (-12,-3)node[anchor=north ]{$1 $} ;
   \draw [thick,rounded corners=4pt] (-11,3)  node[anchor=south]{$2 $   }   -- (-11,1)  -- (-12.5,-.5) -- (-11,-2)   -- (-11,-3)node[anchor=north ]{$2 $}  ;
 \end{tikzpicture}
\end{center}

Note that the Yang-Baxter map \eqref{eq:yba201} is invariant under the diagonal action of $U(n)$, the set of $n\times n$ unitary matrices, on $\CC\PP^{n-1}\times \CC\PP^{n-1}$
and could therefore be defined on $\CC\PP^{n-1}\times \CC\PP^{n-1}/U(n)$. However, this is not the case in general for the reflection map \ref{eq:d13} except 
when $\mm$ is proportional to $I_n$. In this case, the reflection map is proportional to $Id$ when acting on the polarization vectors and thus reduces to the identity map in $\CC\PP^{n-1}$.
Otherwise, in the case \ref{eq:ref22222} with $S$ being a proper subset of $\{1,\dots,N\}$, the action of $U(n)$ results in a different reflection map. This is the mathematical 
translation of the physical effects seen in \cite{2011arXiv1110.2990C} on the polarizations when the so-called boundary basis does not coincide with the polarization basis. 
Therefore, we have found two classes of reflection maps on $\CC\PP^{n-1}$: the identity map and a family parametrized by $U(n)$
\begin{eqnarray}
\rR_U(k):~~\pp\mapsto \left(  I_n + \frac{k-k^*}{k+k^*} \frac{\pp \pp^\dagger}{\pp^\dagger \pp} \right) \,U^\dagger\mathbf{m}U \,\pp\,,
\end{eqnarray}
where
\begin{eqnarray}
\mm=diag(1,\dots,1,-1\dots,-1)\,.
\end{eqnarray}
\section{Reflection maps}
\label{RM}

Let us now present some basic elements of the theory of reflection maps. First we recall some definitions of Yang-Baxter maps.
Let $X$ be a set and ${\cal R} : X \times X \to X \times X$ a map from the Cartesian product of $X$ onto itself. 
Define ${\cal R}_{ij}  : X^N \to X^N$, $X^N = X \times \dots \times X$,  as the map acting as ${\cal R}$ on the $i$th and $j $th factors of the $N$-fold cartesian product $X^N$ 
and identically on the others. More precisely, if $ {\cal R}(x, y) = (f (x, y), g(x, y)) $, $x, y \in X$, then
\begin{align}
i < j\,, \quad           &     {\cal R}_{ij} (x_1, \dots  ,x_n ) = (x_1 , \dots  ,x_{i- 1} , f (x_i , x_j ), \dots ,  g(x_i , x_j ), x_{j +1} , \dots , x_n )\,, \\ 
i > j\,, \quad           &     {\cal R}_{ij} (x_1, \dots  ,x_n ) = (x_1 , \dots  ,x_{i-1} , g (x_i , x_j ) , \dots ,       f(x_i , x_j ), x_{j +1} , \dots , x_n ) \,.
\end{align}
If ${\cal R}_{ij}$ satisfy the following Yang-Baxter relation 
\begin{equation}
  \label{eq:yba112}
    {\cal R}_{12}   {\cal R}_{13}   {\cal R}_{23} = {\cal R}_{23}    {\cal R}_{13}    {\cal R}_{12} \,, 
\end{equation}
then ${\cal R}_{ij}$ is called a Yang-Baxter map.  In particular, for $N = 2$, ${\cal R}_{12} \equiv {\cal R}$.  Let $P$ be the permutation map defined by $P (x, y) = (y, x)$ and 
defined $ {\cal R}_{21} = P {\cal R}_{12}P$. If, ${\cal R}$ satisfies 
\begin{equation}
  \label{eq:yba113}
     {\cal R}_{21}   {\cal R}_{12} = Id\,,
\end{equation}
then ${\cal R}$ is called a reversible Yang-Baxter map. 
       
It is useful to introduce the so-called parametric Yang-Baxter map ${\cal R}_{12}(k_1,k_2)$ which is an important special case obtained by considering 
$X\times Y$ instead of $X$ above, where $Y$ is the set where $k_1$ and $k_2$ live.
It satisfies the parametric YBE
\begin{equation}
  \label{eq:yba114}
   {\cal R}_{12}(k_1, k_2)   {\cal R}_{13}(k_1, k_3)   {\cal R}_{23}(k_2,k_3) = {\cal R}_{23} (k_2,k_3)   {\cal R}_{13} (k_1,k_3)  {\cal R}_{12}(k_1,k_2)\,,
\end{equation}
and the corresponding reversibility condition reads
\begin{equation}
  \label{eq:yba115}
   {\cal R}_{21}(k_2, k_1)   {\cal R} (k_1, k_2) = Id\,.
\end{equation}
We have seen that vector soliton collisions in VNLS provide a solution of such a map. 
Before we introduce a general notion of reflection map, we note that \eqref{eq:d13} requires an extension of the usual definition of parametric maps. In the general acceptance just 
given, the notation ${\cal R}_{12}(k_1,k_2)$ is a short-hand for 
\begin{equation}
{\cal R}_{12}:(x_1,k_1;x_2,k_2)\mapsto (f(x_1,k_1;x_2,k_2),k_1;g(x_1,k_1;x_2,k_2),k_2)\,,
\end{equation}
meaning that the action of ${\cal R}$ on the set $Y$ is trivial. This is not the case for what we defined as a parametric reflection map in \eqref{eq:d13} where we had something of the 
form:
\begin{equation}
 {\rR} : (x, k ) \mapsto   (h(x,k), \sigma(k))\,,
\end{equation}
where $\sigma(k)=-k^*$ was an involution. Defining $S(x,k)=(x,\sigma(k))$ from $X\times Y$ to $X\times Y$, we can reconcile the usual notion of parametric map with our context by 
setting $\rR=SB$ where $B$ is the parametric map $B(x,k)=(h(x,k),k)$. With this definition, the parametric reflection equation \eqref{eq:ref14} should read in fact 
\begin{align}
    S_1 B_1 \yR_{21} S_2B_2 {\cal R}_{12} =\yR_{21} S_2 B_2 \yR _{12} S_1B_1 \,,
\end{align}
as an identity on $(X\times Y)\times (X\times Y)$. We also note that the Yang-Baxter map discussed in this paper has the property
\begin{eqnarray}
S_1S_2\yR_{12}S_1S_2=\yR_{21}\,.
\end{eqnarray}
We now introduce the following general definition.
\begin{definition}
Given four Yang-Baxter maps $\yR^{(j)}$, $j=1,2,3,4$, a reflection map $\rR$ is a solution of the set-theoretical reflection equation
\begin{eqnarray}
\rR_1 \yR^{(2)}_{12} \rR_2 {\cal R}^{(1)}_{12} =\yR_{12}^{(4)} \rR_2 \yR^{(3)}_{12}   \rR_1\,,
\end{eqnarray}
as an identity on $X\times X$.
The reflection map is called involutive if 
\begin{eqnarray}
\rR\rR=Id\,.
\end{eqnarray}
\end{definition}
\begin{remark}
If we restrict our attention to involutive reflection maps and reversible Yang-Baxter maps,  the consistency of the previous definition is ensured 
by requiring that the four Yang-Baxter maps $\yR_{12}^{(j)}$ are related by $\yR_{21}^{(4)}=\yR_{21}^{(2)}=\yR_{12}^{(3)}=\yR_{12}^{(1)}$.
This is assumed in the rest of the paper.
\end{remark}

Allowing for the above extension of the notion of parametric maps, we can define as a special case of the general definition the important class 
of parametric reflection maps.
\begin{definition}
Given the parametric Yang-Baxter map $\yR_{12}(k_1,k_2)$, $k_1,k_2\in Y$ and an involution $\sigma:Y\to Y$, 
a parametric reflection map $\rR(k)$ is a solution of the parametric set-theoretical reflection equation
\begin{equation}
\rR_1(k_1) \yR_{21}(k_1,\sigma(k_2)) \rR_2(k_2) {\cal R}_{12}(k_1,k_2) =\yR_{21}(\sigma(k_1),\sigma(k_2)) \rR_2(k_2) \yR_{12}(\sigma(k_1),k_2)   
\rR_1(k_1)
\end{equation}
as an identity on $X\times X$.
The reflection map is called involutive if 
\begin{eqnarray}
\rR(\sigma(k))\rR(k)=Id\,.
\end{eqnarray}
\end{definition}
Note that in this paper, we explicitely found classes of solution in the case $X=\CC\PP^{n-1}$, $Y=\CC^*$ and $\sigma(k)=-k^*$ with
${\cal R}_{12}(k_1,k_2)$ being the Yang-Baxter map corresponding to VNLS.

We conclude by defining the notion of transfer maps in analogy with those introduced in \cite{veselov2003yang}. Fix $N\ge 2$ and define for 
$j=1,\dots, N$ the following maps of $X^N$ into itself,
\begin{eqnarray}
\label{transfer_map}
{\cal T}_j={\cal R}_{j+1j}\dots {\cal R}_{Nj}\rR_j^-{\cal R}_{jN}\dots{\cal R}_{jj+1}
{\cal R}_{jj-1}\dots{\cal R}_{j1}\rR_j^+{\cal R}_{1j}\dots {\cal R}_{j-1j}\,,
\end{eqnarray}
where $\rR^+$ is a solution of 
\begin{eqnarray}
\label{RE_B+}
\rR_1 \yR_{21}\rR_2 {\cal R}_{12}=\yR_{21} \rR_2 \yR_{12} \rR_1\,,
\end{eqnarray}
and $\rR^-$ a solution of 
\begin{eqnarray}
\rR_1 \yR_{12}\rR_2 {\cal R}_{21}=\yR_{12} \rR_2 \yR_{21} \rR_1\,.
\end{eqnarray}
Then one proves by direct (but long) calculation the following result
\begin{proposition}
For any reversible Yang-Baxter map $\yR$, the transfer maps \eqref{transfer_map} commute with each other
\begin{eqnarray}
{\cal T}_j{\cal T}_\ell={\cal T}_\ell{\cal T}_j~~,~~j,\ell=1,\dots,N\,.
\end{eqnarray}
\end{proposition}

It is known that the set-theoretical YBE has important connection with the braid group (or permutation group in the reversible case) 
acting on $X^N$ \cite{etingof-1997,lu2000set}. Here the formal connection 
of the set-theoretical reflection equation with the finite Artin group (or Weyl group in the involutive case) of type $BC_N$ is quite apparent from \eqref{RE_B+}. 
It is therefore an interesting open question to tackle the construction of general reflection maps in the same spirit as the general construction of Yang-Baxter maps 
in \cite{etingof-1997,lu2000set}. Another interesting avenue for further investigation is that of the role of the transfer maps 
defined above from the point of view of Poisson-Lie groups and integrable dynamics along the lines described in \cite{2006math.....12814V}.

\appendix
\appendixpage
\addappheadtotoc
\printindex

\section{Algorithm for the construction of the "mirror" norming constants}
\label{algo}

We need to solve the coupled nonlinear equations for the $\beta_j$' defined in \eqref{eq:aaa24}. The key observation is relying on Theorem \ref{prop23}. Let
\begin{equation}
  \label{eq:back1111}
  d_{{1}\dots {2N}} (k)= d_{{1}} d_{{2}, \{ {1}\}}  \dots d_{{2N}, \{ {1}\dots  {2N -1} \}} (k)\,. 
\end{equation}
Recall the definition of   ${\cal A}_{j+N}$  in  \eqref{eq:cst3}. The form of the dressing factor \eqref{eq:back1111} leads to the  following expression
\begin{align}
  {\cal A}_{j+N}=  \prod_{i =1, i \neq j+N}^{2N}\left(  \frac{k_{j+N} - k_i}{k_{j+N} - k_i^*} \right) &  \left( d^{-1}_{2N, \{1 \dots 2N-1\}}  \, \dots \, d^{-1}_{j+N+1, \{1 \dots j+N\}} \pi_{j+N, \{1\dots j+N-1\}} \right. \nonumber
\\   \label{eq:rela41} 
&\left.   \, d^{-1}_{j+N-1, \{1\dots j+N-2\}} \, \dots \, d^{-1}_{1} \right) (k_{j+N}) \,,
\end{align}
where
\begin{equation}
  \label{eq:dres145}
  d_{j, \{1\dots j-1\}}(k) = I_n + \left( \frac{k  - k_j}{k - k_j^*} -1 \right)  \pi_{j, \{1\dots j-1\}}\,,
\end{equation}
\begin{equation}
  \label{eq:a14}
 \pi_{j, \{1\dots j-1\}} =    \frac{\xi_{j, \{1\dots j-1\}} \xi^\dagger_{j, \{1\dots j-1\}}}{\xi^\dagger_{j, \{1\dots j-1\}} \xi_{j, \{1\dots j-1\}}}\,, \quad   \xi_{j, \{1\dots j-1\}} =   d^\dagger_{1 \dots j-1} (k_j) \beta_j\,.
\end{equation}
Take $j=N$. Inserting (\ref{eq:rela41}, \ref{eq:dres145}) into  \eqref{eq:aaa24} implies 
\begin{equation}
  \label{eq:rela14}
  \beta_N (\xi_{2N, \{1\dots 2N-1\}})^\dagger =  M(k_N^*)  \prod_{i =1}^{2N-1} \left( \frac{k_{2N} - k_i}{k_{2N} - k_i^*} \right)   \pi_{2N, \{1\dots 2N-1\}}\,.
\end{equation}
Define a $n$-complex vector $\vv_{2N}$ by
\begin{equation}
  \label{eq:real15}
  \vv_{2N} =  \frac{\xi_{2N , \{1\dots 2N-1\}} }{|\xi_{2N, \{1\dots 2N-1\}}|^2} = \left( M(k_N^*)  \prod_{i =1}^{2N-1} \left( \frac{k_{2N} - k_i}{k_{2N} - k_i^*} \right)    \right)^{-1} \beta_N\,. 
\end{equation}
Combining \eqref{eq:rela14} and \eqref{eq:real15} gives 
\begin{equation}
  \label{eq:rela16}
 \xi_{2N , \{1\dots 2N-1\}} = \frac{\vv_{2N}}{|\vv_{2N}|^2}\,.
\end{equation}
With the knowledge of $\xi_{2N , \{1\dots 2N-1\} }$, we can thus compute $d_{2N, \{1 \dots 2N-1\}} (k) $.  Then take $j= N-1$.  From \eqref{eq:a24} and \eqref{eq:rela41}, we  derive
\begin{align}
   \beta_{N-1}  \xi^\dagger_{2N-1, \{1\dots 2N-2\}} =  & M(k_{N-1}^*) \prod_{i =1, i \neq 2N-1}^{2N} \left( \frac{k_{2N-1} - k_i}{k_{2N-1} - k_i^*} \right) \times      \nonumber \\  \label{eq:b11} & \,   d^{-1}_{2N, \{1 \dots 2N-1\}}(k_{2N-1}) \,  \pi_{2N-1, \{1\dots 2N-2\}} \,.
\end{align}
Since  $d_{2N, \{1 \dots 2N-1\}}(k)  $ is known, define $\vv_{2N-1}$ by 
\begin{align}
  \label{eq:b12}
  \displaystyle
  \vv_{2N-1} = & \frac{\xi_{2N -1, \{1\dots 2N-2\}} }{|\xi_{2N-1, \{1\dots 2N-2\}}|^2} \nonumber \\ 
   =& \left( M(k_{N-1}^*)  d^{-1}_{2N, \{1\dots 2N-1\}}(k_{2N-1})\prod_{i =1, i\neq 2N-1}^{2N} \left( \frac{k_{2N-1} - k_i}{k_{2N-1} - k_i^*} \right)    \right)^{-1} \beta_{N-1}\,. 
\end{align}
Combining \eqref{eq:b11} and \eqref{eq:b12} gives 
\begin{equation}
  \label{eq:a1}
  \xi_{2N-1, \{1\dots 2N-2\}}  = \frac{\vv_{2N-1}}{|\vv_{2N-1}|^2}\,.
\end{equation}
We can indeed compute $d_{2N-1, \{1\dots 2N-2\}}(k)$. Recursively, taking $j = N-2$, $N-3$ up to  $ 1$, we are  able to obtain  $\xi_{j+N, \{1 \dots j+N-1\}}$ and  $d_{j+N, \{1 \dots  j+N-1\}}$,  $j = 1, \dots ,N$.  
Since $\{k_j, \beta_j\}_{j\in \{1\dots N\}}$ are known, we have the full knowledge of $d_{j, \{1\dots j-1\}}(k)$, $j =  1, \dots , N$ as well. Therefore, $\beta_{j+N}$ can be derived 
thanks to \eqref{eq:a14}.

\section{Proof of Proposition \ref{prop:refmap1}}
\label{proof_prop_mirror}

To prove \eqref{eq:d11},   we need the mirror symmetry  \eqref{eq:a24} and the permutability property of  dressing transformations (Theorem \ref{prop23}). 
To avoid  tedious notations (the notations are already tiresome!), we choose to  work with  $\{1 \dots  N\}$ instead of $\{i_1 \dots i_N\}$. We 
need to keep in mind that $\{1 \dots  N\}$ can be indeed replaced by any permutation of itself but indexed by the ordered number from $1$ to $N$, 
by taking  $j\to i_j $.  Regardless of permutations,  the relations \eqref{eq:a24} always hold  with respect to the indices.   Write $d_{1\dots 2N}$   
in the following form
\begin{equation}
  \label{eq:a32}
  d_{1\dots 2N}  =    d_{1}   \dots d_{N, \{1\dots N-1\}}  d_{N+1, \{ 1 \dots N\}} \dots   d_{2N, \{1 \dots 2N-1\}}\, 
\end{equation}
Taking  $j = N$ in \eqref{eq:a24},  ${\cal A}_{2N}$ can be written as
\begin{equation}
  \label{eq:a33}
  {\cal A}_{2N} =  \prod_{i =1}^{2N-1}\left(  \frac{k_{2N} - k_i}{k_{2N} - k_i^*} \right)    \pi_{2N, \{1 \dots 2N-1 \}} \, d^{-1}_{1\dots 2N -1}(k_{2N})\,. 
\end{equation}
Substituting this into \eqref{eq:a24} gives
\begin{equation}
  \label{eq:a34}
  \beta_N \xi^\dagger_{2N, \{1 \dots 2N-1 \}} = M(k_N^*)    \prod_{i =1}^{2N-1}\left(  \frac{k_{2N} - k_i}{k_{2N} - k_i^*} \right)    \pi_{2N, \{1 \dots 2N-1 \}} \,.
\end{equation}
Taking the definitions (\ref{eq:fac62}, \ref{eq:facto18}) and  the mirror symmetry $k_{j+N} = -k_j^*$, we come to the following identification
\begin{align}
  \label{eq:a35}
\gamma_{2N, \{1 \dots 2N-1 \}} &=      \xi_{2N, \{1 \dots 2N-1 \}} \,.  \\
  \label{eq:a351}
  \gamma_{N} &=   \prod_{i =1}^{2N-1}\left(  \frac{k_{2N} - k_i}{k_{2N} - k_i^*} \right) ^{-1} \beta_N   \,.  
\end{align}
Define $\uu_{2N}$ by
\begin{equation}
  \label{eq:a36}
  \uu_{2N} =   \frac{ \xi_{2N, \{1 \dots 2N-1 \}} }{ | \xi_{2N, \{1 \dots 2N-1 \}}|^2} = M^{-1}(k^*_N)\,  \gamma_{N}\,.  
\end{equation}
Inserting (\ref{eq:a35} - \ref{eq:a36}) into \ref{eq:a34} gives 
\begin{equation}
  \label{eq:a37}
   \pp_{2N, \{1 \dots 2N-1 \}} =  \frac{\uu_{2N}}{|\uu_{2N}|}=  \mm(k_N) \, \pp_{N}\,.
\end{equation}
Then taking $j=N-1$, we have 
\begin{equation}
  \label{eq:a41}
  {\cal A}_{2N-1} =  \prod_{i =1, i\neq 2N-1}^{2N}\left(   \frac{k_{2N} - k_i}{k_{2N} - k_i^*} \right)    d^{-1}_{2N, \{1\dots 2N-1 \}} \pi_{2N-1, \{1 \dots 2N-2 \}} \, d^{-1}_{1\dots 2N -2}
  (k_{2N-1})\,. 
\end{equation}
Substituting this into \eqref{eq:a24} implies 
\begin{equation}
\label{eq:a42}
d_{2N, \{1\dots 2N-1\}} (k_{2N-1})M^{-1}(k^*_{N-1}) \beta_{N-1} \xi^\dagger_{2N-1, \{1\dots 2N-2\}}   =   \prod_{i =1, i\neq 2N-1}^{2N}\left(   \frac{k_{2N-1} - k_i}{k_{2N-1} - k_i^*} \right) 
 \pi_{2N-1, \{1 \dots 2N-2 \}} \,. 
\end{equation}
With \eqref{eq:a37} and $k_{j+N} = -k_j^*$, we derive 
\begin{align}
  d_{2N, \{1\dots 2N-1\}} (k_{2N-1})M^{-1}(k^*_{N-1})& =   \left(I_n + \left( \frac{k^*_{2N} - k_{2N}}{k_{2N-1} - k^*_{2N}}\right)  \pp_{2N, \{1\dots 2N-1\}}\pp^\dagger_{2N, \{1\dots 2N-1\}} 
   \right) M^{-1}(k^*_{N-1}) \nonumber \\ 
&= M^{-1}(k^*_{N-1} )  \left(I_n +  \frac{k_{N} - k^*_{N}}{k^*_{N-1} - k_{N}}  \pp_{N}\pp^\dagger_{N}  \right)  \nonumber \\
&= M^{-1}(k^*_{N-1} )  d^\dagger_{N}(k_{N-1})\,.
\end{align}
According to (\ref{eq:fac62}), it comes to  the following identification 
\begin{align}
  \gamma_{N-1, \{N\}} =&  \prod^{2N}_{  \substack{
             p=1, p\neq N\\
            p\neq N-1 }}\left( \frac{k^*_{N-1} - k_p}{k^*_{N-1} - k^*_p} \right) \, d^\dagger_{N}(k_{N-1})\,  \beta_{N-1}\,, \\
\gamma_{2N-1, \{1\dots 2N-2\}} =& \left( \frac{k^*_{2N-1} - k_{2N}}{k^*_{2N-1} - k^*_{2N}} \right) \xi_{2N-1, \{1\dots 2N-2\}}\,.
\end{align}
Define 
\begin{equation}
  \label{eq:a38}
  \uu_{2N-1} = \frac{\gamma_{2N-1, \{1\dots 2N-2\}}}{|\gamma_{2N-1, \{1\dots 2N-2\}}|} = M^{-1}(k^*_{N-1}) \gamma_{N-1, \{N\}}\,. 
\end{equation}
Combining (\ref{eq:a42} -  \ref{eq:a38}) together gives
\begin{equation}
  \label{eq:a39}
  \pp_{2N-1, \{1\dots 2N-2\}} = \frac{\uu_{2N-1}}{|\uu_{2N-1}|} = \mm(k_{N-1})\,  \pp_{N-1,\{N\}}\,. 
\end{equation}
Recursively,  taking $j = N-2$, $N-3$ up to $ 1$, the following relations hold
\begin{equation}
  \label{eq:a40}
   d_{q+N, \{1\dots q-1+N\}}(k_{j+N})\, M^{-1}(k^*_{j}) =  M^{-1}(k^*_{j}) \, d^\dagger_{q, \{q+1 \dots N\}}(k_j) \,, \quad j \le q\, . 
\end{equation}
Then  inserting $k_{j+N} = -k_{j}^*$, $  \gamma_{j, \{j+1\dots N\}} $ and $\gamma_{j+N, \{1\dots j+N\}} $  into  \eqref{eq:a24} yields  
\begin{equation}
  \label{eq:d12}
    \pp_{{j+N}, \{j\dots {j-1+N}\}} =  \mm(k_{j}) \,   \pp_{j,  \{{j+1} \dots N\}}\,.  
\end{equation}
By applying the correspondence between $\{1, \dots , N\}$ and   $\{i_1, \dots, i_N\}$ with  $j \to i_j$, we come to \eqref{eq:d11}.
 Equation (\ref{eq:c11} is obtained in the same way by assuming the following form of the dressing factor
\begin{align}
  \label{eq:a21}
  &d_{1\dots 2N}  =    d_{1}   \dots d_{j-1, \{1 \dots j-2\}}  d_{j+1, \{ 1 \dots j-1\}} \dots   d_{N, \{1 \dots\hat{j}\dots N-1\}}  d_{j, \{1 \dots\hat{j}\dots N\} }  
 d_{j+N, \{1 \dots N\}} \nonumber \\ & \dots  d_{j+N-1, \{1 \dots j+N-2\}}  d_{j+N+1, \{1 \dots j+N-1\}} \dots d_{2N, \{1 \dots\widehat{j+N}\dots 2N-1\} } 
 d_{j+N , \{1 \dots\widehat{j+N}\dots 2N\}}\,. 
\end{align}
This  means that $  d_{j, \{1 \dots\hat{j}\dots N\}} $ is the last dressing factor added in the product of the first $N$ dressing factors, and   $  d_{j+N, \{1 \dots\widehat{j+N}\dots 2N \}} $ 
is the last dressing factor added in the product of the total $2N$ dressing factors. Then applying  \eqref{eq:d11} yields directly (\ref{eq:c11}).


\end{document}